\documentclass[nd]{iosart2x}

\usepackage{subcaption}

\def\ojoin{\setbox0=\hbox{$\bowtie$}%
  \rule[-.02ex]{.25em}{.4pt}\llap{\rule[\ht0]{.25em}{.4pt}}}
\def\leftouterjoin{\mathbin{\ojoin\mkern-5.8mu\bowtie}}

\newcommand{\IMPL}		{\ensuremath{\rightsquigarrow}}

\newcommand{\obl}	{\ensuremath{\mathbf{O}}}

\newcommand{\compl}{\ensuremath{\rhd}}	

\newcommand{\actionsp}   {\ensuremath{\mathcal{AP}}}

\newcommand{\deontics}  {\ensuremath{\mathcal{D}}}

\newcommand{\rules}     {\ensuremath{\mathcal{R}}}

\newcommand{\extendedActions}   {\ensuremath{\mathcal{EAP}}}

\newcommand{\extendedObligations}   {\ensuremath{\mathcal{EO}}}

\newcommand{\violatedO}   {\ensuremath{\mathcal{VO}}}
\newcommand{\activeO}   {\ensuremath{\mathcal{AO}}}
\newcommand{\fulfilledO}   {\ensuremath{\mathcal{FO}}}
\newcommand{\notSatisifedO}   {\ensuremath{\mathcal{NSO}}}
\newcommand{\expiredO}   {\ensuremath{\mathcal{EO}}}

\usepackage{listings}
\usepackage[english]{babel}
\usepackage{amsthm}
\usepackage{natbib}  
\usepackage{comment}
\theoremstyle{definition}
\newtheorem{definition}{Definition}
\usepackage[linesnumbered,ruled,vlined]{algorithm2e}
\usepackage{float} 
\usepackage{enumerate}
\usepackage{multirow}

\pubyear{0000}
\volume{0}
\firstpage{1}
\lastpage{1}

\begin{document}

\begin{frontmatter}

\title{Modeling and Managing Temporal Obligations in GUCON Using
SPARQL-star and RDF-star}
\runtitle{Managing Temporal Obligations in GUCON}


\begin{aug}
\author[A]{\inits{N.}\fnms{Ines} \snm{Akaichi}\ead[label=e1]{ines.akaichi@wu.ac.at}%
\thanks{Corresponding author. \printead{e1}.}}
\author[B]{\inits{N.N.}\fnms{Giorgos} \snm{Flouris}\ead[label=e2]{fgeo@ics.forth.gr}}
\author[B]{\inits{N.-N.}\fnms{Irini} \snm{Fundulaki}\ead[label=e3]{fundul@ics.forth.gr}}
\author[A]{\inits{N.-N.}\fnms{Sabrina} \snm{Kirrane}\ead[label=e4]{sabrina.kirrane@wu.ac.at}}
\address[A]{Institute for Complex Networks, \orgname{Vienna University of Economics and Business,} \cny{Austria}\printead[presep={\\}]{e1,e4}}
\address[B]{Institute of Computer Science, \orgname{FORTH}, Heraklion, \cny{Greece}\printead[presep={\\}]{e2,e3}}
\end{aug}


\begin{abstract}
In the digital age, data frequently crosses organizational and jurisdictional boundaries, making effective governance essential. Usage control policies have emerged as a key paradigm for regulating data usage, safeguarding privacy, protecting intellectual property, and ensuring compliance with regulations. A central mechanism for usage control is the handling of obligations, which arise as a side effect of using and sharing data.
Effective monitoring of obligations requires capturing usage traces and accounting for
temporal aspects such as start times and deadlines, as obligations may evolve over times into different states, such as fulfilled, violated, or expired. While several solutions have been proposed for obligation monitoring, they often lack formal semantics or provide limited support for reasoning over obligation states. To address these limitations, we extend GUCON, a policy framework grounded in the formal semantics of SPAQRL graph patterns, to explicitly model the temporal aspects of an obligation. This extension enables the expressing of temporal obligations and supports continuous monitoring of their evolving states based on usage traces stored in temporal knowledge graphs.
We demonstrate how this extended model can be represented using RDF-star and SPARQL-star and propose an Obligation State Manager that monitors obligation states and assess their compliance with respect to usage traces. Finally, we evaluate both the extended model and its prototype implementation. 
\end{abstract}


\end{frontmatter}




\section{Introduction}


In modern decentralized environments, such as data spaces and knowledge graph applications, data routinely moves across organizational and jurisdictional boundaries. The unrestricted flow of information brings with it the need for continuous and context-aware enforcement of policies to ensure the governance of data usage. In this context, usage control has emerged as a paradigm for safeguarding privacy, protecting intellectual property rights, and ensuring compliance with regulations \cite{AKAICHI2025100698}. Unlike traditional access control, which primarily focuses on determining who is permitted to access a given resource, usage control extends this notion by regulating not only access but also how, when, and under what conditions the data may subsequently be used, modified, or shared \cite{Sandhu2004}. 

A central mechanism for operationalization these conditions and ensuring continuous policy enforcement is the concept of obligations \cite{Hilty2007}.  Obligations specify the required actions that must be performed before, during, or after data usage thereby ensuring accountability, compliance, and trust throughout the entire data usage process. 
Examples of such obligations include the requirement to delete data within a period of five years, or to notify the data owner whenever a document is shared with third parties. For this, effective monitoring of obligations requires the incorporation of temporal aspects such as the start time and deadline, as obligations may evolve through different states, such as fulfilled, violated, or expired \cite{Fornara2018}, as time passes.
The enforcement of obligations follows two main strategies \cite{DBLP:journals/ieeesp/PretschnerHSSW08}: preventive and detective. Preventive mechanisms delay an attempted usage request until the corresponding obligations have been fulfilled, ensuring that actions are taken before access or further processing occurs. In contrast, detective mechanisms focus on verifying compliance after an obligation has been executed, typically through auditing or logging tools. The latter depends upon a representation of the state of affairs that captures domain knowledge and records events in the form of logs or usage traces \cite{akaichi2023}. As highlighted by \citet{Hilty2007}, many obligations are difficult to enforce proactively; for example, confirming the permanent deletion of data is challenging to ensure in real time but can be assessed retrospectively through log records.  

Several models have been proposed for usage control and obligation monitoring in particular. The Usage CONtrol model (UCON) \cite{Sandhu2004} introduced obligations, decision continuity, and attribute mutability, yet despite various formalizations and integration attempts \cite{LAZOUSKI201081, Maurizio2010}, no standard specification or widely adopted implementation exists.
Policy languages such as the Obligation Specification Language (OSL) \cite{Hilty2007}, Rei \cite{Kagal2002}, Ponder \cite{Damianou2001}, Proteus \cite{Toninelli2005},  KAoS \cite{Uszok2003}, and  the Dynamic Spectrum Access Policy Framework (DSA policy framework) \cite{10.1007/978-3-030-62466-8_30} support temporal constraints alongside obligations but lack models for obligation states, limiting their ability to track obligation life cycles. The Open Digital Rights Language (ODRL) \cite{ODRL2018}, while widely recognized, lacks official formal semantics, hindering rigorous reasoning about temporal aspects such as obligation start times and deadlines. Furthermore, the majority of the proposed solutions do not model the state of affairs, an essential component for storing usage traces and enabling dynamic reasoning over obligation monitoring and compliance with usage policies in general.

In order to address these gaps, we initially proposed the Generic \underline{G}raph Pattern-based Policy Framework for \underline{U}sage \underline{Con}trol Enforcement (GUCON) \cite{akaichi2023}, which is a usage control framework that allows permission, prohibition, obligation, and dispensation policies to be expressed. We also define the notion of the state of the affairs, which is a knowledge graph that captures domain knowledge and events, and serves as the basis for reasoning about and enforcing usage control policies. 
In this work, we extend the GUCON framework by incorporating temporal properties into obligations, enabling their monitoring over time and demonstrate how it can be instantiated using RDF-star and SPARQL-star. Our main contributions are as follows: (i) we extend the original GUCON obligation model with start times and deadlines, allowing reasoning about obligation states;
(ii) we formally define two key reasoning tasks that determine the states of temporal obligations (such as active, fulfilled, or violated) and assess the compliance of the knowledge base with respect to these defined obligations;  
(iii) we demonstrate how this extended model can be represented using RDF-star and SPARQL-star and propose an Obligation State Manager that monitors obligation states and checks compliance of obligations against the state of affairs; and (iv) we evaluate the extended model and its prototype implementation.

The remainder of this paper is structured as follows: Section \ref{sec:usecase} presents a motivating use case describing the process of inpatient care in a hospital. Section \ref{sec:prelim} introduces the preliminaries in relation to RDF, SPARQL, and  GUCON. Section \ref{sec:semantics} defines the formal semantics of the extended GUCON model in order to cater for temporal obligations. Section \ref{sec:implementation} describes our RDF-star, SPARQL-star, and Obligation State Manager instantiations and Section \ref{sec:evaluation} evaluates both the extended obligation model and our Obligation State Manager. Section \ref{sec:relatedwork} subsequently positions our work with respect to the state of the art. Finally, Section \ref{sec:conclusion} concludes and discusses future work.


\section{Use Case}
\label{sec:usecase}

\begin{figure}[t] 
  \centering
  \includegraphics[width=0.85\textwidth]{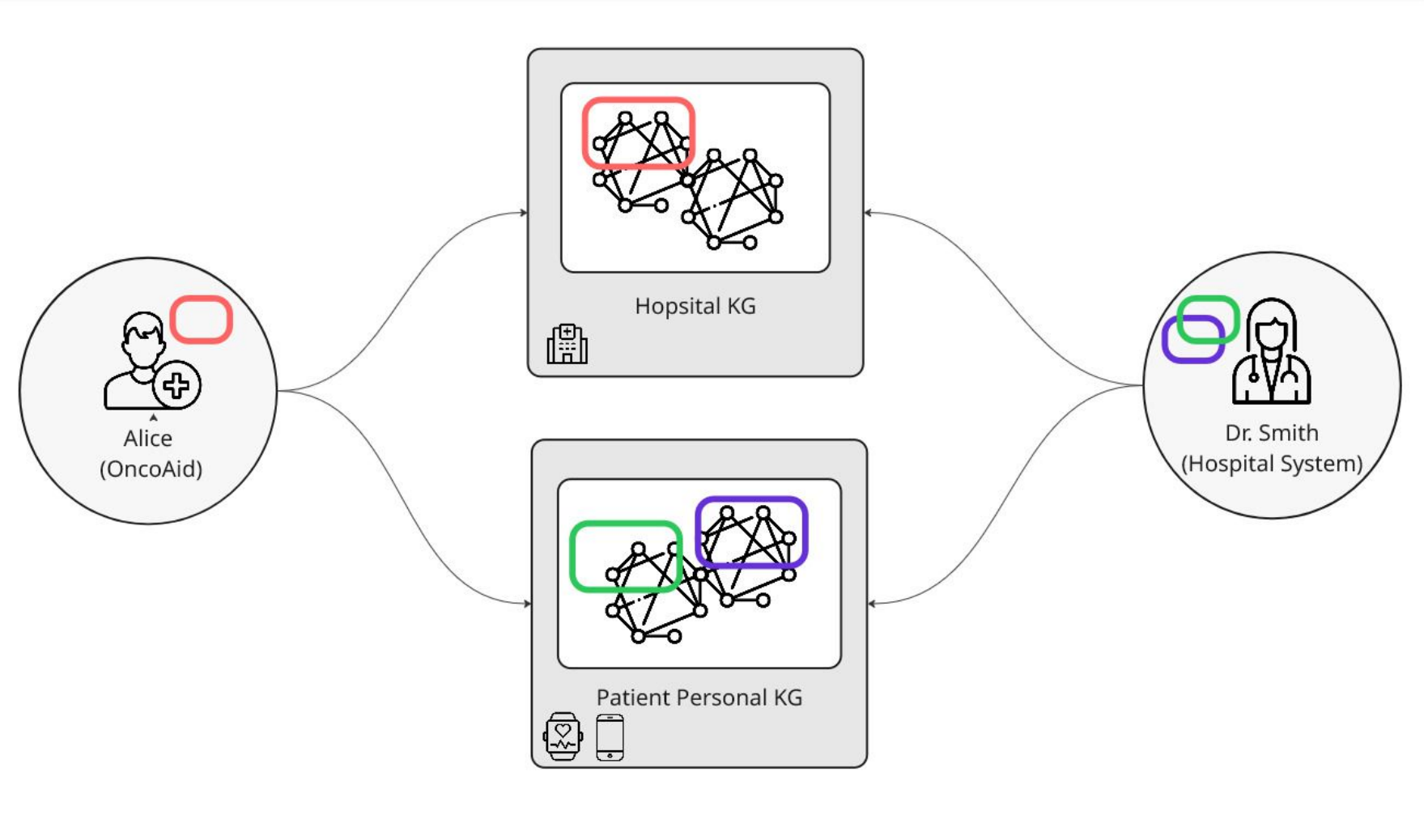}
  \caption{A Depiction of the OncoAid Use Case}
    \label{fig:use-case}
\end{figure}
Consider an imaginary application, OncoAid, which is designed to support cancer patient health monitoring by enabling them to manage their medical and health-related data. The app is integrated with a smartwatch that continuously collects and monitors vital signs (e.g., heart rate and blood pressure), which are stored in the patient's personal Knowledge Graph (KG).
In addition to real-time monitoring, OncoAid provides access to key medical data, including electronic medical records, lab results, imaging, medication plans, and doctor diagnostic notes. This data is maintained in the hospital's KG.
Patients using OncoAid can view and manage their personal data through their own KG while also accessing relevant data from the hospital KG. While, doctors are responsible for managing hospital data and, when necessary, accessing the patient’s KG to provide informed medical care.
\smallskip

\noindent\textbf{Running Example.} The running example, which is depicted in Figure \ref{fig:use-case}, illustrates the process of inpatient care at CityCare Hospital, where Alice is admitted for treatment. Upon admission, her corresponding doctor, Dr. Smith, performs different laboratory tests.
By the end of her stay, a diagnosis is concluded, and all diagnosis notes and laboratory results are stored in the hospital KG. During Alice's hospitalization, we envisage the following scenarios:



\begin{itemize}
    \item[] \textbf{Scenario 1 (S1).} 
    During Alice's stay, Dr. Smith orders lab tests, in which the results are stored in the hospital KG. Dr. Smith \emph{must} devise a treatment plan and share it with Alice after her lab results are ready. 
    \item[] \textbf{Scenario 2 (S2).}
    As Alice’s expected discharge date approaches, she \emph{must} review and electronically sign her discharge form before the scheduled end of her inpatient care. This electronically signed document is securely stored in the hospital’s KG.
    \item[] \textbf{Scenario 3 (S3).} 
    After Alice is discharged from CityCare Hospital, her doctor  \emph{must} sign and finalize her diagnostic report within 12 hours. Alice's discharge from CityCare marks the actual end of her inpatient care.
\end{itemize}
The different scenarios highlight various obligations governing data usage. Each obligation is regulated by temporal rules that determine whether an action must occur before, after, or within a specified time window.

\section{Preliminaries}
\label{sec:prelim}
The Generic \underline{G}raph Pattern-based Policy Framework for \underline{U}sage \underline{Con}trol Enforcement (GUCON) \cite{akaichi2023} provides an abstract and semantically grounded structure to specify usage control policies that support their implementation. Additionally, the framework defines approaches for policy reasoning tasks, including consistency, requirements, and compliance checking\footnote{In this paper, we focus primarily on compliance and requirements checking, as our emphasis is solely on obligations.}.
These  tasks are defined based on the concepts of state of affairs and usage control policies, which rely on RDF and SPARQL.  Accordingly, the formal basis for reasoning is provided by SPARQL semantics.
In the following, we begin by presenting the essential SPARQL preliminaries, followed by an introduction to GUCON.

\subsection{RDF \& SPARQL}
In our work, we rely on the syntax and semantics of graph patterns expressions presented in \cite{Perez2006}.
Throughout the paper, we assume two pairwise disjoint and infinite sets $I$ and $L$ to denote respectively Internationalized Resource Identifiers (IRIs) and literals. We denote by $IL$ the union of $I \cup L$. Note that blank nodes are omitted for simplicity.
We introduce the concepts of subject \texttt{s}, property \texttt{p}, and object \texttt{o} to form \texttt{subject}-\texttt{property}-\texttt{object} RDF triples. An RDF triple $tr = (s,p,o) \in TR$ , where  $TR = (s,p,o) \in (I) \times (I) \times (I\cup L)$. A set of RDF triples forms an \textit{RDF graph} $D$.
We assume additionally the existence of an infinite set $V$ of variables disjoint from the above sets. As a notational convention, we will prefix variables with ``?''(e.g., ?x, ?y).

\paragraph{Syntax of Graph Patterns.}
The definition of graph patterns is based on triple patterns. A triple pattern is defined as $(sp,pp,op) \in$ $(I \cup V) \times (I \cup V) \times (I\cup L \cup V)$. The variables occurring in a graph pattern $G$ are denoted as var($G$).
\begin{definition} [Graph Pattern]
\label{def:graph-pattern}
A graph pattern is defined recursively as follows:
\begin{itemize}
    \item A triple pattern is a graph pattern.
    \item If $G1$ and $G2$ are graph patterns, then (G1 AND G2), (G1 OPT G2), (G1 UNION G1), (G1 MINUS G2) are graph patterns.
    \item If G is a graph pattern and R is a filter expression, then (G FILTER R) is a graph pattern. A filter expression is constructed using elements of the sets $I \cup L \cup V$, logical connectives ($\neg, \wedge, \vee$), inequality symbols ($<, \leq,\geq, >$), equality symbol ($=$), plus other features (see \cite{Prudhommeaux2006} for a complete list).
\end{itemize}
\end{definition}
\paragraph{Semantics of Graph Patterns.}
The semantics of graph pattern expressions are defined based on a partial mapping function $\mu$, where $\mu: V \rightarrow TR $. For a triple pattern $tp$, we denote by $\mu(tp)$ the triple obtained by replacing the variables in $tp$ according to $\mu$. The domain of $\mu$, $dom(\mu)$, is the subset of $V$ where $\mu$ is defined. 
\begin{definition}[Evaluation of a Graph Pattern]
\label{def:evaluation}
 Let $D$ be an RDF graph over $TR$. Mapping a graph pattern against $D$ is defined using the function $[[.]]_{D}$, which takes a graph pattern expression and returns a set of mappings $\Omega$.   
\end{definition}
Two mappings $\mu_1$ and $\mu_2$ are said to be compatible, denoted by $\mu_1 \sim \mu_2$ when, for all $x \in dom(\mu_1) \cap dom(\mu_2)$, it is the case that $\mu_1(x)=\mu_2(x)$. The evaluation of a compound graph pattern is defined as follows:\\
$[[G_1$ AND $G_2]]$ = $\Omega_1 \bowtie \Omega_2 = \{ \mu_1 \cup \mu_2 \,|\, \mu_1 \in \Omega_1, \mu_2 \in \Omega_2$, $ \mu_1 \sim \mu_2$\}\\
$[[G_1$ UNION $G_2]]$ = $\Omega_1 \cup \Omega_2 = \{ \mu_1 \cup \mu_2 \,|\, \mu_1 \in \Omega_1 \cup \mu_2 \in \Omega_2$, $ \mu_1 \sim \mu_2$\}\\
$[[G_1$ OPT $G_2]]$ = $\Omega_1 \leftouterjoin \Omega_2 = (\Omega_1 \bowtie \Omega_2) \cup (\Omega_1/\Omega_2)$, where\\
$\Omega_1 / \Omega_2 = \{ \mu | \mu \in \Omega_1 \wedge \nexists \mu^\prime \in \Omega_2, \mu \sim \mu^\prime \}$

\subsection{GUCON}

The GUCON framework is composed of two core components: the state of affairs and the usage control policies. The state of affairs is represented by an RDF graph, whereas the usage control policies consist of rules that are defined based on SPARQL graph patterns. 
These components are fundamental to the definition and implementation of the reasoning tasks of the framework, which are based on the formal semantics of SPARQL. 
In what follows, we first present the syntax of the state of affairs and the usage control policies (with a focus on obligations), followed by a formal definition of their semantics and the associated reasoning tasks.

\subsubsection{Specification}
The state of affairs captures both the relevant domain knowledge and observed events, serving as the foundation for evaluating and enforcing usage control policies. Throughout this paper, the state of affairs will be referred to as the knowledge base (KB).

\begin{definition} [Knowledge Base]
A KB, denoted as $K$, is an RDF graph describing the set of actual knowledge. 
\end{definition}

Obligations guide the behavior of entities in their use of resources by specifying the actions they are required to perform under certain conditions. For instance, Scenario \textbf{S3} specifies \emph{an obligation for Dr. Smith (entity) to sign (action) Alice’s diagnosis report (resource)}. 

We assume three sets, $N$ (entity names), $C$ (action names), and $R$ (resource names), such that $N$, $C$, $R \subseteq I$. We define an \textit{action} as a special RDF triple $(n,c,r) \in (N) \times (C) \times (R)$.
In practice, we will allow variables to be present in the $n$, $c$, $r$ positions (e.g., \texttt{?doctor ?sign ?diagnosisReport}), to allow generally applicable restrictions to be expressed. We refer to such a tuple as an \textit{action pattern}. 

\begin{definition} [Action Pattern]
An action pattern is a triple $(np,cp,rp) \in (N \cup V) \times (C \cup V) \times (R \cup V)$. We denote by \actionsp\ the set of all action patterns.
\end{definition}

An obligation pattern can be defined as follows.

\begin{definition} [Obligation Pattern]
\label{def:deontic-action-pattern}
Let $ \obl$ denotes an \emph{Obligation}.
An \emph{obligation pattern} is a statement of the form $\obl a$, where $a \in \actionsp$.
\end{definition}
In the context of an action $a$ defined as $(n, c, r)$, $\obl (n, c, r)$: indicates that an entity $n$ is obliged to perform an action $c$ over a resource $r$.
An obligation usually applies under specific conditions, giving rise to conditional obligation  rules (e.g, \emph{the obligation for Dr. Smith to sign a Alice’s diagnosis report applies only under certain conditions. Dr. Smith must be a doctor, Alice must be a patient, and a diagnosis report for Alice must already exist, etc.}).
Using an obligation pattern, a graph pattern describing the \emph{condition} of the rule, and the operator $\IMPL$ in-between, a conditional deontic rule, simply called an obligation rule, can be defined as follows. 
\begin{definition} [Obligation Rule]
\label{def:usage-control-rule}
An obligation rule is 
of the form: $cond \IMPL \obl a$,
where $cond$ is a graph pattern, and $ \obl a $ is an obligation pattern.
We denote by $\rules$ the set of all obligation rules. For each rule, it is required that $var(a) \subseteq var(cond)$.
\end{definition}
A obligation rule can be read as follows: If the condition ($cond$) is satisfied by the KB, then the obligation pattern ($\obl a$) must also hold. The restriction $var(a) \subseteq var(cond)$ ensures that all variables appearing in the obligation are already bound by the condition. Without this restriction, the model would implicitly generate infinitely many rules, which would render it impractical.
An obligation policy consists of a collection of such obligation rules.
\begin{definition} [Obligation Policy]
\label{def:policy}
A set of obligation rules $R \subseteq \mathcal{R}$ is called an obligation policy. 
\end{definition}

\subsubsection{Reasoning over GUCON Policies}
To effectively perform the reasoning tasks in our framework, it is important that we have the ability to reason about the rules governing obligation policies. 
This is primarily accomplished through the process of evaluating these rules against the KB, resulting in the identification of \textit{active rules}. Active rules are characterized by having a \textit{satisfied condition}, ensuring their applicability in the given KB.
\begin{definition} [Satisfied Condition]
\label{def:satis-condition}
Let K\ be a KB, $R$ an obligation policy, and $r=cond \IMPL \obl a \in R$ an obligation rule. 
A condition $cond$ is satisfied for $\mu$, denoted by $K \compl cond$, if and only if 
there exists a mapping $\mu$ such that  $\mu \in [[cond]]_{K}$.
\end{definition}
Note that multiple mappings may  be used to satisfy a given rule.
\begin{definition} [Active Obligation Rule]
\label{def:active-rule}
An  obligation rule $r =cond \IMPL \obl a \in R$ is active for a mapping $\mu$, if and only if $cond$ is satisfied for $\mu \in [[cond]]_K$. 
\end{definition}
Note that a rule may be active for multiple mappings. 
Based on the definition of an active rule, a usage control policy $R$ is called \textit{active} if at least one of its rules is active. Otherwise, it is called \textit{inactive}.

Based on the defined formal semantics, different reasoning tasks are envisioned such as compliance and requirements checking\footnote{Note that these tasks were defined in the original GUCON paper in the context of a specific entity $n$. For the sake of generalization and simplicity, we redefine these tasks in a general sense, independent of any entity.}. 
Compliance checking aims to identify any breaches or non-compliant behavior of an entity, thereby ensuring the proper and secure operation of the system. In the context of a KB, compliance checking involves verifying that the stored facts and usage traces adhere to predefined policies.
The criteria for determining whether a KB is compliant with a given policy can vary depending on the specific rule being considered, as well as the KB and mappings used to interpret that rule. 
\begin{definition} [KB Compliance Against an Obligation]
\label{def:compliance-rule}
Given a KB $K$ and an obligation rule $r = cond \IMPL \obl a$, we say that $K$ complies with $r$, denoted by $K \compl r$, if and only if $\mu(a) \in K$ whenever $\mu \in [[cond]]_K$.
\end{definition}
On the other hand, requirements checking is a task that enables a system to query a policy and receive information regarding the set of active obligations given a KB $K$, denoted by $R_{active}$.  
\begin{definition} [Requirements Checking]
\label{def:requirement-checking}
Given a KB $K$ and a set of rules $R$, the set of active rules $R_{active}$ is defined as follows: $R_{active} = \{\mu(r)| r\in R $,  r is active for $\mu$\}.
\end{definition}
Intuitively, $R_{active}$ contains the concrete rule instances that are triggered by the current state of the KB, obtained by instantiating the general rules 
$R$ with substitutions that satisfy their conditions.

\section{Formal Semantics of Temporal GUCON Obligations}
\label{sec:semantics}
While GUCON enables the specification of various data usage conditions through graph patterns, incorporating constraints like temporal restrictions requires extending the rule’s deontic pattern. This extension  allows for the expression of obligations with a temporal dimension, such as deleting data within 30 days of a deletion request or after a period of five years from the time of acquisition.
Building upon previous work in temporal knowledge representation \cite{10.1007/11431053_7}, we extend the GUCON model of obligation rules to encompass temporal properties. Consequently, the KB is also extended to represent these temporal attributes.

\subsection{Temporal Knowledge Base}
As part of our extension, we include the notion of an event  that presents an action executed at a specific time. 
\paragraph {Event \& Event Pattern.} When an action $(n,c,r)$ is executed at a time $t_{exec}$, then an event $(n,c,r,t_{exec})$ is created in the KB. An event is a tuple $(n, c, r, t_{exec})  \in  (N  \times C  \times R  \times  T )$. Similarly, an event pattern is a tuple $(np, cp, rp, tp_{exec})  \in (N \cup V) \times (C \cup V) \times (R \cup V) \times (T \cup  V) 
$.

Our KB store two kinds of triples: (1) facts that do not change over time. For example, \emph{Dr. Smith is a doctor}. Although facts may evolve over time, for simplicity, we  exclude their temporal aspects. These facts are stored in the Data Knowledge Base (DKB), which is represented as an RDF graph.
(2) Events that capture actions occurring at specific points in time, such as \emph{a doctor shares a patient record at time $t$}. These are stored in the Event Knowledge Base (EKB), represented as a temporal graph. Both the DKB and the EKB are part of the KB.
Building on the definition of temporal graphs proposed by \citet{10.1007/11431053_7}, an EKB is defined as follows. 
\begin{definition} [Event Knowledge Base] 
An event is a tuple $(n,c,r,t_{exec})$. An EKB is a finite set of events.
\end{definition}

Given an EKB $K^e$, we can define a snapshot of $K^e$ at time t. 

\begin{definition} [Event Knowledge Base Snapshot]
Given a time t, a snapshot of an EKB, denoted as $K_{t}^{e}$, includes only events that happened before t:
$K_{t}^{e} = \{ (n,c,r)|\exists t_{exec}, (n,c,r,t_{exec})\in K^e$ and $ t_{exec} <= t \} $. 
\end{definition}
Intuitively, an EKB snapshot at time $t$ captures the state of the KB by including only those events that have already occurred up to $t$.
Given a DKB $K^s$ and a snapshot of an EKB $K^e_{t}$, the full snapshot of the KB at t is the union: $K_{t} = K^s \cup K_{t}^{e} $.

\subsection{Temporal Obligation Rules}
To enable GUCON obligations to account for temporal properties, we enhance the deontic actions within the rules by incorporating temporal attributes into both actions and action patterns.

\paragraph{Extended Action \& Action Pattern.}
We extend the action pattern of an obligation with meta-properties describing the start time and deadline of the obligation. The start time of an obligation $t_{start}$ presents the time when the obligation starts being applicable, whereas the deadline $t_{deadline}$ presents the time by which the obligation must be fulfilled.  
Scenario \textbf{S3} illustrates \emph{the obligation for Dr. Smith to sign Alice’s diagnosis report within 12 hours of her actual hospitalization end date}. The start time refers to the actual hospitalization end date, and the deadline falls exactly 12 hours after.
In some cases, $t_{start}$ or $t_{deadline}$ can be undefined. 
However, both  $t_{start}$ and $t_{deadline}$ cannot be undefined at the same time.
When $t_{start}$ is undefined; the obligation expresses an action that needs to be executed before a certain deadline but without any specific starting point; in other words, an obligation that is in force until it expires at its deadline. Scenario \textbf{S2} illustrates \emph{Alice's obligation to sign a discharge form before the scheduled end of her hospitalization}.
The undefined $t_{start}$ is represented using the symbol $-\infty$ .
When $t_{deadline}$ is undefined, i.e., when the obligation has no expiration date, the obligation expresses an action that needs to be executed after a certain point of time marked by the start time. Scenario \textbf{S1} shows \emph{the doctor's obligation to share a treatment with Alice after her lab results are ready}.
In this case, the undefined $t_{deadline}$ is represented using the symbol $\infty$.
We assume the set $T$ presenting all literals that represent time, including $\infty$ and $-\infty$. 
An extended action is a tuple $(n, c, r, t_{start},t_{deadline})  \in  (N  \times C  \times R  \times  T 
\times  T )$;  with the additional constraint: $(t_{start} = -\infty )  \oplus  (t_{deadline} = \infty)$, i.e., either $t_{start}$ or $t_{deadline}$ are defined as specific time points.

\begin{definition} [Extended Action Pattern]
An extended  action pattern is a tuple $(np, cp, rp, tp_{start},tp_{deadline})  \in (N \cup V) \times (C \cup V) \times (R \cup V) \times (T \cup  V) 
\times  (T \cup V) $. We denote by {\extendedActions} the set of all extended action patterns. 
\end{definition}

Given the extended action patterns, an obligation pattern is also extended as follows.
\begin{definition} [Extended Obligation Pattern]
\label{def:deontic-action-pattern-star}
An \emph{extended obligation pattern} is of the form $\obl {ea}$, where $\obl \in \deontics$ and $ea \in \extendedActions$.
\end{definition}
Similarly, an obligation rule is extended as follows.
\begin{definition} [Extended Obligation Rule]
\label{def:usage-control-rule-star}
An extended obligation rule is 
of the form: $cond \IMPL \obl {ea}$,
where $cond$ is a graph pattern, and $\obl {ea}$ is an extended deontic pattern. We denote by $\extendedObligations$ the set of all extended obligation rules.    
\end{definition}

\subsection{Reasoning Tasks}
Building upon the extension of GUCON to incorporate temporal properties, the different  reasoning tasks such as requirements and compliance checking can be further refined. In particular, requirements checking can address queries like identifying  fulfilled, expired, or violated obligations at a time $t$, commonly referred to as obligation states \cite{Fornara2019}.  

\begin{figure}[t] 
  \centering
  \includegraphics[width=0.38\textwidth]{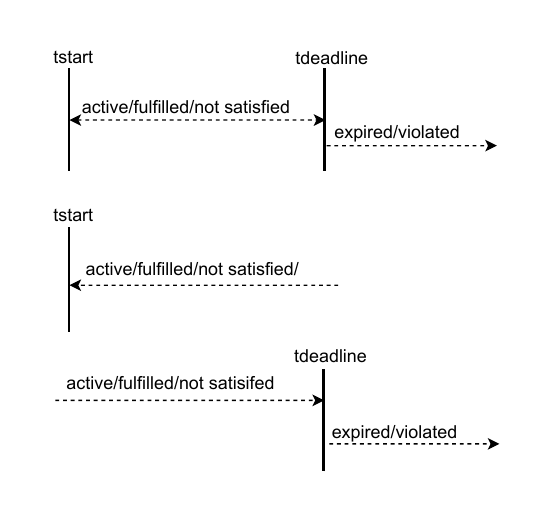}
  \caption{States Of Obligations.}
   \label{fig:states}
\end{figure}

Generally, given a snapshot  $K_{t}$ of a KB at time $t$, an extended obligation rule $OR$  of the general form $ cond \IMPL \obl (np, cp, rp, tp_{start},tp_{deadline})$, $OR$ can either be \emph{active}, \emph{fulfilled}, \emph{violated}, \emph{expired}, or \emph{not satisfied}. Figure \ref{fig:states} illustrates the states of obligations with regards to the start time and/or deadline.
The different states of obligations depend on the identification of rules such as the condition is satisfied given a snapshot of a KB  at a time t. 
We extend the definition of satisfied condition by including the time aspect as follows.
\begin{definition} [Satisfied condition]
\label{def:satis-condition-1}
A condition $cond$ is satisfied for $\mu$ given a snapshot $K_{t}$, denoted by $K_{t} \compl cond$, if and only if 
there exists a mapping $\mu$ such that  $\mu \in [[cond]]_{K_{t}}$.
\end{definition}
An active obligation is an obligation whose condition is satisfied and  has not yet expired, i.e., it remains enforceable for a specified period. If the start time or deadline is undefined, the obligation still qualifies as active as long as at least one of these time points is specified.

\begin{definition} [Active Obligation]
\label{def:active-obligation}
An obligation $OR$ is active for a mapping $\mu$ at time $t$, if and only if $cond$ is satisfied for $\mu$ at $t$ and $\mu(tp_{start})<=t<=\mu(tp_{deadline})$.  
We denote by {\activeO}  the set of active obligations at a time $t$.
\end{definition}

A fulfilled obligation requires the corresponding action to be executed within the designated time frame. If either the start time or the deadline is undefined, the obligation is still considered fulfilled as long as at least one of these time points is specified.


\begin{definition} [Fulfilled Obligation]
\label{def:fulfilled-obligation}
An obligation $OR$ is fulfilled at time $t$ if only if $cond$ is satisfied for a mapping $\mu$ at $t$, $\exists t_{exec}$, such that $(\mu(np,cp,rp),t_{exec}) \in K_{t}$ and $ 
\mu(tp_{start})<=t_{exec}<=\mu(tp_{deadline})$. 
We denote by $\fulfilledO$ the set of all fulfilled obligations at time $t$. 
\end{definition}

A violation of obligation represents an action that was not performed while the deadline has passed. In case the deadline is undefined, an obligation is never violated.

\begin{definition} [Violated Obligation]
\label{def:violated-obligation}
An obligation $OR$ is violated at time $t$ if and only if $cond$ is satisfied for a mapping $\mu$ at $t$, $\nexists t_{exec}$, such as 
$ (\mu(np,cp,rp),t_{exec})  \in K_{t}$ and  $t>\mu(tp_{deadline}) $.  
We denote by \violatedO the set of violated obligations at a time $t$. 
    
\end{definition}

An expired obligation is an obligation that is no longer enforceable passed a specific deadline. In case the deadline is not defined, an obligation would never expire.  


\begin{definition} [Expired Obligation]
\label{def:expired-obligation}
An obligation $OR$ is expired at time $t$ if and only if $cond$ is satisfied for a mapping $\mu$ at $t$ and $t>\mu(tp_{deadline})$. 
We denote by {\expiredO} the set of expired obligations at time $t$. 

\end{definition}

A not satisifed obligation is an obligation that has not yet been fulfilled, but the deadline has not passed. If either the start time or the deadline is undefined, the obligation is still considered not satisfied as long as at least one of these time points is specified.

\begin{definition} [Not Satisfied Obligation]
\label{def:notSatisfied-obligation}
An obligation $OR$ is not satisfied at time t if only if $OR$ is active for a mapping $\mu$,  $\nexists t_{exec}$, such as 
$ (\mu(np,cp,rp),t_{exec})  \in K_{t}$, and $ \mu(tp_{start})<=t<=\mu(tp_{deadline})$. 
We denote by {\notSatisifedO} the set of not satisifed obligations at time $t$. 
\end{definition}
Compliance checking evaluates whether the KB reflects the possible states of obligations. A KB is compliant if no expired obligations are shown as violated.
\begin{definition} [Compliant Knowledge Base at time t]
\label{def:compli-kb}
A KB $K$ is compliant at time t if and only if:
$\expiredO \cap \violatedO = \emptyset$.
\end{definition}

\section{Implementation: An obligation Manager for Temporal GUCON Obligations}
\label{sec:implementation}
In this section, we present the syntax used to express temporal GUCON. We also provide a proof of concept implementation that supports the reasoning tasks described in Section \ref{sec:semantics}, along with the ontologies employed to support the proposed solution\footnote{Throughout the paper, we use the prefixes \emph{gucon},  \emph{hc}, and \emph{ucp} to denote respectively the GUCON core, the Hospital Inpatient Care (HIC) and the Usage Control Policy (UCP) vocabularies.}.

\subsection{Implementing Temporal GUCON using RDF-star and SPARQL-star}
In what follows, we present an overview of SPARQL-star and RDF-star. Then, we demonstrate how to represent our GUCON temporal obligations using SPARQL-star and RDF-star syntax.


\subsubsection{RDF-star and SPARQL-star}

The RDF-star data model extends RDF by allowing
arbitrarily deep nesting of triples as subject or object arguments.  Any RDF triple $tr \in TR$ is an RDF-star triple. Given RDF-star triples $tr$ and $tr'$, and RDF terms $s \in I $, $p \in I$,
and $o \in (I \cup L)$, then the triples $(tr,p,o)$, $(s,p,tr)$ and $(tr,p,tr')$ are also RDF-star triples.

\begin{definition}[SPARQL-star Triple Pattern]
\label{def:tripleStar}
A SPARQL-star triple pattern is a 3-tuple that is defined recursively as follows:
\begin{itemize}
\item  Every SPARQL triple pattern is a SPARQL-star triple pattern;
\item  If tr and tr' are SPARQL-star triple patterns, $x \in (IL \cup V) $, $p \in (I \cup V)$, then $(tr, p, x)$, $(x, p, tr)$, and $(tr, p, tr')$ are SPARQL-star triple patterns. 
\end{itemize}
\end{definition}

\begin{definition}[SPARQL-star Graph Pattern]
A SPARQL-star graph pattern is defined recursively as follows:
\label{def:graphStar} 
\begin{itemize}
\item A SPARQL-star triple pattern is a SPARQL-star graph pattern.
\item If $G1$ and $G2$ are SPARQL-star graph pattern, then (G1 AND G2), (G1 OPT G2), (G1 UNION G1), (G1 MINUS G2) are SPARQL-star graph pattern.
\item If G is a SPARQL-star graph pattern and R is a filter expression, then (G FILTER R) is a SPARQL-star graph pattern. A Filter expression is constructed using elements of the sets $I \cup L \cup V$, logical connectives ($\neg, \wedge, \vee$), inequality symbols ($<, \leq,\geq, >$), equality symbol ($=$), plus other features (see \cite{Prudhommeaux2006} for a complete list).
\end{itemize}

\end{definition}
The semantics of SPARQL-star graph pattern expressions are similar to the semantics of standard SPARQL graph patterns \cite{Perez2006}.
The notion of a SPARQL-star solution mapping extends the notion of a standard SPARQL solution mapping; that is, every SPARQL solution mapping is a SPARQL-star solution mapping. However, in contrast to SPARQL solution mappings, SPARQL-star solution mappings may map variables also to RDF-star triples.

{\footnotesize
\begin{algorithm}[t]
\DontPrintSemicolon
\SetAlgoLined
\SetKwInOut{Input}{Input}
\SetKwInOut{Output}{Output}
\SetKwFunction{getObligationsStates}{GET\_OBLIGATIONS\_STATES}
\SetKwFunction{GetSnapshot}{GET\_SNAPSHOT}
\SetKwFunction{GetMappings}{GET\_MAPPINGS}
\caption{Evaluate States of Policy Rules at Time $t$}
\label{algo:states}
\Input{Policy $P$, Knowledge Base $K$, Time $t$}
\Output{Object \texttt{states}}

\SetKwProg{Fn}{Function}{:}{}
\Fn{\getObligationsStates{$P$, K, t}}{

$K_t \leftarrow$ \GetSnapshot{$K$, $t$} \;
\texttt{states} $\leftarrow$ new \texttt{ObligationStates}() \;

\ForEach{rule $r \in P$}{
    $\omega \leftarrow$ \GetMappings{$r.condition$, $K_t$} \;
    \If{$\omega \ne \emptyset$}{
        \ForEach{binding $\mu \in \omega$}{
            mappedRule $\leftarrow \mu(r)$ \;

            \If{mappedRule.startTime $\leq t \leq$ mappedRule.deadline}{
              \texttt{states.activeO} $\leftarrow$ \texttt{states.activeO} $\cup$ \{mappedRule\}
              
                \If{mappedRule.executionTime $=$ \texttt{null}}{
                    \texttt{states.notSatisifedO} $\leftarrow$ \texttt{states.notSatisifedO} $\cup$ \{mappedRule\} \;
                }
              
            }
            \ElseIf{$t >$ mappedRule.deadline}{

                \texttt{states.expiredO} $\leftarrow$ \texttt{states.expiredO} $\cup$ \{mappedRule\} \;
                
                \If{mappedRule.executionTime $=$ \texttt{null} \textbf{or} mappedRule.executionTime $>$ mappedRule.deadline}{
                    \texttt{states.violatedO} $\leftarrow$ \texttt{states.violatedO} $\cup$ \{mappedRule\} \;
                }
         
            }
            \ElseIf{mappedRule.executionTime $\ne$ \texttt{null}}{
                \If{mappedRule.startTime $\leq$ mappedRule.executionTime $\leq$ mappedRule.deadline}{
                    \texttt{states.fulfilledO} $\leftarrow$ \texttt{states.fulfilledO} $\cup$ \{mappedRule\} \;
                }
            }
        }
    }
}
\Return{\texttt{states}} \;
}
\end{algorithm}

}

\subsubsection{Syntax for Temporal GUCON }
In order to represent the temporal aspects of GUCON rules and the temporal KB, we adopt SPARQL-star and RDF-star that enable the use of embedded triples to annotate statements with temporal metadata. 

\paragraph{Temporal Obligation Rules.}  
Temporal rules are enriched with the temporal predicates \emph{gucon:startTime} and \emph{gucon:deadline}, which respectively denote the start time and deadline associated with an obligation. An extended action pattern is expressed as a SPARQL-star triple pattern\footnote{RDF-star and SPARQL-star, \url{https://w3c.github.io/rdf-star/cg-spec/2021-07-01.html}}:
\[
\ll ?n\ ?a\ ?r \gg\ \emph{gucon:startTime}\ tp_{\text{start}} ;\ \emph{gucon:deadline}\ tp_{\text{deadline}} .
\]
where $?n\ ?a\ ?r$ represents an embedded triple pattern and
$tp_{\text{start}}$, $tp_{\text{deadline}}$ map to literals of type \emph{xsd:dateTime}.
Rule conditions may also contain SPARQL-star triple pattern, making both the action pattern and its condition SPARQL-star graph patterns.

\paragraph{Temporal Knowledge Base.}  
Temporal facts in the KB are modeled using the predicate \emph{gucon:executionTime}, which links an executed action to its \emph{xsd:dateTime} value. 
{\footnotesize
\begin{algorithm}[t]
\DontPrintSemicolon
\SetAlgoLined
\SetKwInOut{Input}{Input}
\SetKwInOut{Output}{Output}
\SetKwFunction{getObligationsStates}{GET\_OBLIGATIONS\_STATES}
\SetKwFunction{CheckCompliance}{CHECK\_COMPLIANCE}

\caption{Check Compliance of a Knowledge Base at Time $t$}
\label{algo:compliance}

\Input{Policy $P$, Knowledge Base $K$, Time $t$}
\Output{Compliance status}
\SetKwProg{Fn}{Function}{:}{}
\Fn{\CheckCompliance{$P$, K, t}}{
    \texttt{states} $\leftarrow \getObligationsStates{P, K, t}$ \;

    \If{$\texttt{states.expiredO} \cap \texttt{states.violatedO} \ne \emptyset$}{
        \Return{\texttt{NON\_COMPLIANT}} \;
    }

    \Return{\texttt{COMPLIANT}} \;
}
\end{algorithm}

}
This is expressed as:
\[
\ll n\ a\ r \gg\ \emph{gucon:executionTime}\ t_{\text{exec}} .
\]
Where $t_{\text{exec}}$ is a value of type \emph{xsd:dateTime}. The KB is thus an RDF-star graph, capable of expressing both static and temporal information.

\paragraph{Rule Evaluation with Temporal Events.} Given a rule of the form:
\[
\emph{cond} \ \IMPL\ \obl \left\{ \ll ?n\ ?a\ ?r \gg\ \emph{gucon:startTime}\ tp_{\emph{start}} ;\ \emph{gucon:deadline}\ tp_{\text{deadline}} \right\} .
\]
To retrieve possible execution times from the KB, the rule is \emph{augmented} with an optional temporal binding as follows:
\[
\begin{aligned}
&\emph{cond} \ \texttt{OPTIONAL} \left\{ \ll ?n\ ?a\ ?r \gg\ \emph{gucon:executionTime}\ tp_{\text{exec}} \right\} \\
&\IMPL \obl \left\{ \ll ?n\ ?a\ ?r \gg\ \emph{gucon:startTime}\ tp_{\text{start}} ;\ \emph{gucon:deadline}\ tp_{\text{deadline}} \right\}.
\end{aligned}
\]
The use of \texttt{OPTIONAL} ensures that if an execution time exists in the KB for the action pattern, it is retrieved; otherwise, the rule can still be processed using only the other bindings. This enriched rule is then matched against the RDF-star KB, allowing the values for start time, deadline, and (if available) execution time to be extracted and used in further reasoning tasks. Listing \ref{lst:oblig1} shows an instantiation of Scenario \textbf{S3} using SPARQL-star. While \emph{?startTime} and \emph{?deadline} are not always explicitly provided, they can be computed dynamically using the \texttt{BIND} operator. These values are eventually derived from the actual start time stored in the KB and the duration specified in the policy. Scenarios \textbf{S1} and \textbf{S2} can be expressed in a similar way, using one of the temporal predicates.
 
\subsection{System Architecture}

In order to show a proof of concept of our defined semantics for temporal GUCON obligations, we developed a GUCON Obligation State Manager that implements the reasoning tasks defined in Section \ref{sec:semantics}. 
The GUCON Obligation Manager is implemented in Java using Apache Jena 5.1.0. The manager expects three primary inputs: a KB and a policy, both provided as Turtle files, and a time instant $t$, expressed using the W3C XML Schema Definition Language (XSD)\footnote{W3C XML Schema Definition Language (XSD), \url{https://www.w3.org/TR/xmlschema11-2/}}.

The Obligation Manager comprises five core components: the \emph{Knowledge Base Manager}, the \emph{Rule Manager}, the \emph{Obligation State Manager}, the \emph{Compliance Checker}, and the \emph{Report Generator}. The Knowledge Base Manager loads the KB from a Turtle file into Jena TDB2. The Rule Manager loads and maintains the rules in memory as Java objects, enabling their evaluation against the KB. Additionally, it is responsible for augmenting the rules with optional temporal bindings. 
The Obligation State Manager is responsible for reasoning over the states of obligations, while the Compliance Checker assesses the compliance of the KB. The Report Generator produces a compliance report detailing the state of each obligation and the overall compliance status.

The core logic of the Obligation State Manager and the Compliance Checker is implemented through the following key algorithms, which expect as input a GUCON policy $P$, a KB $K$ and a time $t$. 
Algorithm~\ref{algo:states} handles the management of obligation states in three main steps: i) It extracts a snapshot $Kt$ of the KB $K$ at time $t$ (line 2); ii) For each rule, it determines whether the rule is satisfied in $Kt$ by invoking the \texttt{GET\_MAPPINGS} function, which evaluates the rule over the snapshot and returns a set of mappings that populate the rule with concrete values (lines 5–8); iii) Finally, the algorithm evaluates the state of each mapped obligation based on its start time, deadline, and the input time $t$ (lines 9–25). For simplicity, we illustrate only the case in which both the start time and deadline are defined; for the other scenarios, the same reasoning is followed.
While Algorithm~\ref{algo:compliance} is responsible for checking the compliance of the KB. It assumes that the status of each rule has already been determined by invoking the function in Algorithm~\ref{algo:states} (line 2). If the sets of violated and expired rules are disjoint (line 3), then $K$ is considered compliant. The final output indicates the compliance status of the input KB at a given time $t$.

\subsection{Ontology-Driven Representation for the GUCON Obligation State Manager}
In the following, we present the OWL-based vocabularies used to model the input and output of our Obligation State Manager. We also provide an RDF instantiation of our use case, demonstrating how the defined vocabularies are applied in practice.

\subsubsection{Core Ontologies Supporting GUCON}
To streamline the management of inputs and outputs within the GUCON Obligation Manager, we developed dedicated ontologies that encode both the GUCON Policy and the Compliance Report. In contrast, the KB integrates ontologies specific to the application domain, with a particular focus on vocabularies for temporal representation.

\begin{figure}[t] 
 \centering
 \includegraphics[width=0.80\textwidth]{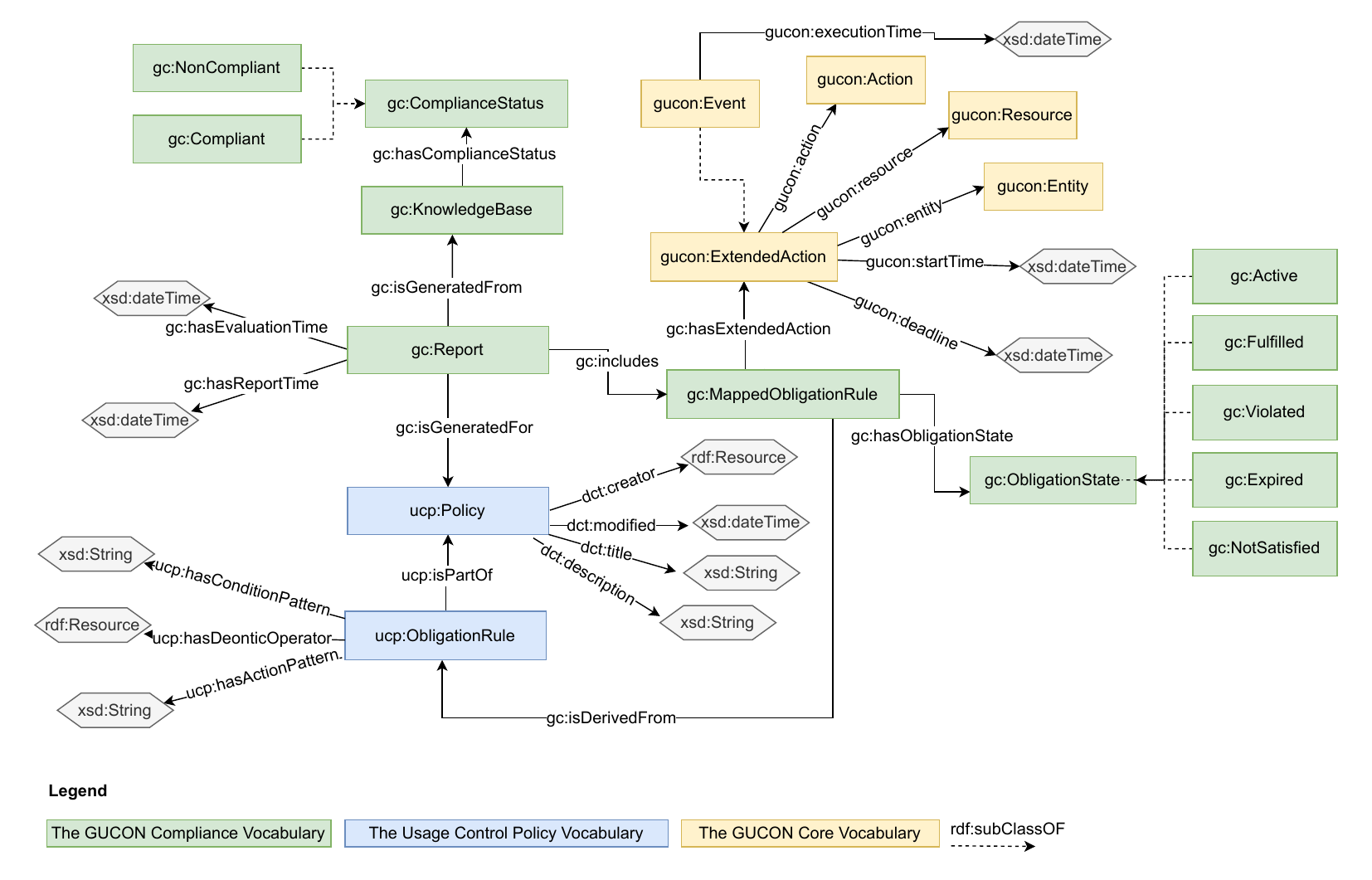}
 \caption{An ontology for a Compliance Report}
  \label{fig:report}
\end{figure}

\paragraph{An RDF Vocabulary for GUCON Policies.} To ensure interoperability and facilitate automated processing, GUCON policies are represented in RDF. Figure~\ref{fig:report} (highlighted in blue) presents the Usage Control Policy (UCP) vocabulary, which defines GUCON obligation rules through the properties \emph{ucp:hasConditionPattern}, \emph{ucp:hasActionPattern}, and \emph{ucp:hasDeonticOperator}. The property \emph{ucp:isPartOfPolicy} indicates the policy to which a given rule belongs.  Furthermore, instances of \emph{ucp:Policy} can be enriched with metadata, such as creation date, creator, and description, by reusing properties from the Data Catalog Vocabulary (DCAT)\footnote{Data Catalog Vocabulary (DCAT), \url{https://www.w3.org/TR/vocab-dcat-3/}}.

\paragraph{An RDF Vocabulary for Compliance Reports.} Figure~\ref{fig:report} shows the ontology depicting a compliance report.
A compliance report provides a detailed description of the states of mapped rules and the compliance status of the KB. The class \emph{gc:Report} is generated for a \emph{ucp:Policy} and associated with a \emph{gc:KnowledgeBase}. The evaluation time of the report indicates the time $t$ at which a snapshot of the input KB is taken, whereas the report time refers to the actual moment the report is generated.
The compliance status of the KB is represented by the \emph{gc:ComplianceStatus} class. The ontology also captures the relationship between mapped obligation rules and the original input rules. A \emph{gc:MappedObligationRule} represents an obligation rule that has been instantiated with specific data from the KB. It is linked to a \emph{gucon:ExtendedAction}, which is composed entirely of constants. The state of each mapped rule is captured using the \emph{gc:ObligationState} class.
An \emph{gucon:ExtendedAction}, connects to an \emph{gucon:Entity}, an \emph{gucon:Action}, and a \emph{gucon:Resource}. This action is further characterized by the temporal properties \emph{gucon:startTime} and \emph{gucon:deadline}, indicating the start and deadline of the associated obligation. Additionally, the property \emph{gucon:executionTime} is used to specify the exact time at which an event occurs. When an extended action is executed, it is represented as an instance of the \emph{gucon:Event} class.
The various ontologies can be found on our Github\footnote{\url{https://github.com/Ines-Akaichi/Temporal-GUCON/tree/main/ontologies}}.

\subsubsection{Use Case Instantiation}
To illustrate the  practical applicability of the proposed GUCON Compliance ontology and the Obligation State Manager, we model the use case described in Section \ref{sec:usecase} using the Hospital Inpatient Care (HIC) ontology\footnote{The HIC ontology is  accessible on our GitHub repository, \url{https://github.com/Ines-Akaichi/Temporal-GUCON/tree/main/ontologies/domain}}.
Specifically, we demonstrate how the GUCON State Manager processes input data and generates compliance outcomes based on defined policies and observed actions with a focus on  Scenario \textbf{S3}, as it involves both a start time and a deadline.  Listing~\ref{lst:oblig1-policy} shows the RDF encoding of Scenario \textbf{S3} following the UCP vocabulary.

\noindent
\begin{minipage}[t]{0.48\textwidth}
\lstset{ basicstyle=\scriptsize, frame=single}
\begin{lstlisting} [
basicstyle=\scriptsize, %or \small or \footnotesize etc.
caption={Scenario 3 Modeled as a GUCON Rule},captionpos=b, label={lst:oblig1}, numbers=left,]
{?doctor a hc:Doctor .
?diagnosisReport a hc:DiagnosisReport .
?admission a hc:Admission .
?admission hc:hasPatient ?patient .
?admission hc:hasActualAdmissionEndDate 
    ?actualAdmissionEndDate .
?diagnosisReport hc:hasAdmission  ?admission .
?patient hc:hasResponsibleDoctor ?doctor .
BIND (?actualAdmissionEndDate AS ?startTime) .
BIND(?startTime +"PT12H"^^xsd:duration AS ?deadline)}
-> 
O 
{
<<?doctor gucon:sign ?diagnosisReport >> 
gucon:startTime ?startTime; gucon:deadline ?deadline.
}
\end{lstlisting}
\end{minipage}
\hfill
\begin{minipage}[t]{0.48\textwidth}
\lstset{ basicstyle=\ttfamily\small, frame=single}
\begin{lstlisting} [
basicstyle=\scriptsize, %or \small or \footnotesize etc.
caption={RDF-Based Encoding of Scenario 3},captionpos=b, label={lst:oblig1-policy}, numbers=left,]
exp:rule-obligation-sign-diagnosis-report 
a ucp:ObligationRule;
ucp:hasActionPattern 
    "<<?doctor gucon:sign ?report>> ..."
ucp:hasConditionPattern 
    "?patient rdf:type  hc:Patient ...";
ucp:hasDeonticOperator  
    ucp:ObligationRule;
ucp:isPartOfPolicy      
    exp:policy-obligation-sign-diagnosis-report.
exp:policy-obligation-sign-diagnosis-report   
    a ucp:Policy ;
    dcat:description "an example policy" ;
    dcat:creator  ex:ines-akaichi ;
    dcat:modified  
    "2025-07-20T10:30:00+02:00"^^xsd:dateTime .                           
\end{lstlisting}
\end{minipage}

Listing \ref{lst:kb} provides a corresponding snapshot of the KB. We assume that when this KB is loaded into the system, it is assigned a unique IRI \emph{ex:kb-trace-doctor-angelika-smith} for identification.
The KB shows that the condition (lines 1-10) in Listing \ref{lst:oblig1} is satisfied. Notably, the KB also stores the execution of the required action, signing the diagnosis report, using the predicate \emph{gucon:hasExecutionTime}.


\begin{lstlisting}[
basicstyle=\scriptsize, %or \small or \footnotesize etc.
caption={Scenario 3 Corresponding Knowledge Base},captionpos=b, label={lst:kb}, numbers=left, frame=single]
ex:doctor-angelika-smith a hc:Doctor ;
    hc:name "Angelika Smith" .
 ex:patient-alice-waltz a hc:Patient ;
    hc:name "Alice Waltz" ;
    hc:hasResponsibleDoctor  ex:doctor-angelika-smith .
ex:admission-alice-waltz-2025-07-15 a hc:Admission ;
    hc:hasActualAdmissionStartDate "2025-07-15T09:30:00+02:00"^^xsd:dateTime ;
    hc:hasActualAdmissionEndDate "2025-07-20T10:30:00+02:00"^^xsd:dateTime;
    hc:hasPatient ex:patient-alice-waltz .
ex:diagnosis-report-alice-waltz-2025-07-15  a hc:DiagnosisReport ;
    hc:hasAdmission ex:admission-alice-waltz-2025-07-15 .
<< ex:doctor-angelika-smith  gucon:sign ex:diagnosis-report-alice-waltz-2025-07-15 >>
    gucon:executionTime  "2025-07-20T12:30:00+02:00"^^xsd:dateTime; 
\end{lstlisting}

Based on the policy definition in Listing~\ref{lst:oblig1-policy} and the KB content in Listing~\ref{lst:kb}, Listing~\ref{lst:compliance} shows an excerpt from the resulting compliance report.
The report shows that the mapped obligation rule \emph{instance-obligation-sing-diagnosis-form-01} is classified as both fulfilled and expired. It further specifies that the obligation’s start time corresponds to the patient’s hospitalization end date, as stated in the main policy rule, while the evaluation time occurs after the rule’s deadline.
At the time of evaluation, the KB confirms that the doctor  \emph{ex:kb-trace-doctor-angelika-smith} completed the required action, signing the diagnosis report, within the permissible time window defined by the obligation’s start time and deadline.
Therefore, the KB \emph{exp:kb-trace-doctor-angelika-smith} is deemed fully compliant at the time of policy evaluation.

\begin{lstlisting} [
basicstyle=\scriptsize, %or \small or \footnotesize etc.
caption={Scenario 3 Corresponding Compliance Report},captionpos=b, label={lst:compliance}, numbers=left,frame=single]
exp:instance/obligation-sign-diagnosis-report-01     rdf:type   ucp:MappedObligationRule;
    gucon:hasExtendedAction    exp:sign-alice-waltz-sign-diagnosis-report-2025-07-20  ;
    gucon:isDerivedFrom        exp:rule-obligation-sign-diagnosis-report  ;
    gucon:hasObligationState   gucon:FULFILLED, gucon:EXPIRED .
exp:sign-alice-waltz-sign-diagnosis-report-2025-07-20   rdf:type   gucon:Event ;
     gucon:hasEntity  ex:doctor-angelika-smith;
     gucon:hasAction <gucon:sign> ;
     gucon:hasResource  ex:diagnosis-report-alice-waltz-2025-07-15 ;
     gucon:hasStartTime  "2025-07-20T10:30:00+02:00"^^xsd:dateTime;
     gucon:hasDeadline   "2025-07-20T22:30:00+02:00"^^xsd:dateTime ;
     gucon:hasExecutionTime  "2025-07-20T12:30:00+02:00"^^xsd:dateTime .
exp:report-2025-07-20T11:01:00+02:00
    gucon:hasEvaluationTime  "2025-07-21T10:00:00+02:00"^^xsd:dateTime;
    gucon:hasReportTime   "2025-07-21T10:00:12:00+02:00"^^xsd:dateTime;
    gucon:isGeneratedFor     exp:policy-obligation-sign-diagnosis-report;
    gucon:isGeneratedFrom    exp:kb-trace-doctor-angelika-smith;
    gucon:includes           exp:instance-obligation-sign-diagnosis-report-01 .
exp:kb-trace-doctor-angelika-smith    gucon:hasComplianceStatus     gucon:COMPLIANT .
\end{lstlisting}



\section{Evaluation}
\label{sec:evaluation}
This section presents our evaluation strategy, which comprises two complementary assessment components: the GUCON model and the GUCON State Manager. The GUCON model is evaluated in accordance with Thörn’s quality assessment criteria \cite{thorn2010quality}, providing a structured analysis of its conceptual and design quality. In parallel, the GUCON State Manager is assessed using a benchmark design inspired by the methodology developed within the HOBBIT 2020 project\cite{doi:10.3233/DS-190021}. For this second evaluation, we also describe the experimental setup used to conduct the benchmark and present the corresponding results.

\subsection{Thörn’s Criteria for Model Quality}
Inspired by previous work that  applied Thörn's quality assessment criteria \cite{thorn2010quality} to evaluate policy models \cite{10.1007/978-3-030-62466-8_30}, we use the criteria to assess our extended GUCON model.
Thörn’s Criteria for model quality provide a systematic approach for evaluating the quality of modeled artifacts and encompass the following factors: \emph{changeability}, \emph{reusability}, \emph{formalness}, \emph{correctness}, \emph{mobility}, and \emph{usability}. 

\emph{Changeability.} This criteria illustrates the model’s ability to evolve alongside its applications while preserving its core purpose. A fundamental aspect of usage control is the capability to express and reason about temporal aspects, enabling the monitoring of obligations \cite{Sandhu2004}. Other constraints, such as spatial, purpose-based, and event-driven restrictions are also important for capturing realistic usage control scenarios \cite{Hilty2007}. We have first addressed the temporal dimension, demonstrating how GUCON can represent and reason about temporal constraints. The same approach can be extended to incorporate other constraint types, such as spatial restrictions, by enriching the extended obligation action pattern with additional meta-properties. For example, spatial attributes can be captured through latitude and longitude values, resulting in an enhanced extended action pattern of the form:
$(np, cp, rp, tp_{start},tp_{deadline}, tp_{latitude}, tp_{longitude })$, which expresses an action that must be executed within a specific time frame at a certain location. 
SPARQL-star is a particularly suitable option for encoding this extended action pattern, as it offers a concise representation without excessive verbosity.

\emph{Reusability.} This refers to the ability to reuse the model (or parts of the model) across different use cases or applications. GUCON rules are designed to be general and adaptable, requiring only domain-specific information to represent particular scenarios. Our extended GUCON obligation model captures the core elements of an obligation: the entity, the resource, the actions, and their associated temporal aspects. By combining the model with a suitable ontology or vocabulary, it can be tailored to represent specific use cases.
A broad spectrum of usage control scenarios can be envisioned, spanning domains such as healthcare, mobility, energy, and agriculture \cite{AKAICHI2025100698}. In our work, we developed the HIC ontology to model the medical use case described in Section~\ref{sec:usecase}. Both the ontology and the represented scenarios are publicly accessible in our GitHub repository\footnote{\url{https://github.com/Ines-Akaichi/Temporal-GUCON/tree/main/use-case}}. Other ontologies could similarly be employed to capture domain knowledge in mobility \cite{KATSUMI201853}, energy \cite{BOOSHEHRI2021100074}, or agriculture \cite{10.1007/978-3-642-18333-1_18}.

\emph{Formalness.} This refers to the model's ability to be defined and managed in a formalized  way. In our case, the GUCON model is grounded in the formal semantics of SPARQL graph patterns, which are based on model-theoretic semantics. This foundation facilitates the precise specification and management of the model's semantics, leveraging the well-established formal semantics underpinning SPARQL \cite{Perez2006}.
Our model employs operators such as \texttt{AND} and \texttt{UNION} to represent conjunctive and disjunctive conditions. The \texttt{OPT} operator functions similarly to the outer join in SQL, while \texttt{MINUS} allows the expression of negation in conditions (e.g., to identify all patients who did not sign the discharge form, one could write \emph{?patient rdf:type  hc:Patient  \texttt{MINUS}  ?patient  gucon:sign  ?dischargeForm}. The \texttt{FILTER} operator is used to specify conditions on particular sub-elements of a triple. Additionally, assignment properties such as \texttt{BIND} can be used to compute derived values, particularly date-time values. For instance, in Scenario \textbf{S3}, we demonstrated how \texttt{BIND} can be used to calculate the deadline of an obligation.

\emph{Correctness.} This factor concerns the model’s effectiveness in capturing and addressing real-world requirements, particularly through the representation of domain-specific artifacts. We employ the developed HIC ontology to represent the use case scenarios described in Section~\ref{sec:usecase}. Furthermore, to assess the model’s expected output with respect to obligation states, we design test cases for each scenario. In each case, the expected states are determined based on the start time, deadline, and execution time of the corresponding obligation, together with the given input time $t$. Table~\ref{tab:correctness} presents the developed test cases together with their expected states. The corresponding KBs and rules for each scenario are made available on our GitHub\footnote{https://github.com/Ines-Akaichi/Temporal-GUCON/tree/main/test-cases}.

\begin{table}[t]
\centering
\caption{Test Cases Representing the Hospital Inpatient Care Scenarios}
\label{tab:correctness}
\begin{tabular}{lllllll}
\textbf{Sceanrio} & \textbf{Case} & \textbf{Start Time} & \textbf{Deadline} & \textbf{Time} & \textbf{Execution Time} & \textbf{Expected States} \\
\multirow{2}{*}{\textbf{S1}} & S11 & yes & no  & after or equal to start time & none                         & active, not satisfied \\
                             & S12 & yes & no  & after start time             & after or equal to start time & active, fulfilled     \\
                             \hline
\multirow{4}{*}{\textbf{S2}} & S21 & no  & yes & before or equal to deadline  & none                         & active, not satisfied \\
                             & S22 & no  & yes & before or equal to deadline  & before or equal to deadline  & active, fulfilled     \\
                             & S23 & no  & yes & after deadline               & none                         & expired, violated     \\
                             & S24 & no  & yes & after deadline               & before or equal to deadline  & expired, fulfilled     \\
                             \hline 
\multirow{4}{*}{\textbf{S3}} & S31 & yes & yes & during interval              & none                         & active, not satisfied \\
                             & S32 & yes & yes & during interval              & during interval              & active, fulfilled     \\
                             & S33 & yes & yes & after deadline               & none                         & expired, violated     \\
                             & S34 & yes & yes & after deadline               & during interval              & expired, fulfilled   \\
                             \hline
\end{tabular}
\end{table}


\emph{Mobility.}  This criteria addresses the model’s interoperability and its potential for integration with other systems. In this paper, we demonstrated how the extended GUCON model can be represented using the SPARQL-star syntax. Beyond SPARQL-star, our model can also be instantiated with other widely adopted W3C languages, such as the Shapes Constraint Language (SHACL)\footnote{SHACL, \url{https://www.w3.org/TR/shacl/}} and the Web Ontology Language (OWL)\footnote{OWL, \url{https://www.w3.org/TR/owl2-primer/}}.
SHACL was originally designed for validating RDF graphs, but it has also attracted attention for expressing policies \cite{Robaldo2021,robaldo2023compliance}, particularly through its advanced features\footnote{SHACL advanced features, \url{https://www.w3.org/TR/shacl-af/}}. For example, a usage control obligation can be represented in SHACL as a \texttt{sh:SPARQLRule}, where the \texttt{sh:construct} property defines a construct query. Using this representation, the template clause specifies the action pattern, while the \texttt{where} clause can be directly mapped to the condition of the obligation.
Similarly, OWL has been extensively used to express policies in the Semantic Web \cite{Kagal2002,10.1007/978-3-030-62466-8_30}. In OWL, an obligation can capture the entity, resource, and action using \texttt{owl:equivalentClass} together with \texttt{owl:intersectionOf} (to represent conjunctions, i.e., logical AND). These elements of a rule  can be recursively described using \texttt{owl:hasValue} or \texttt{owl:allValuesFrom}. The logical \texttt{UNION} operator can be expressed via \texttt{owl:unionOf}, while \texttt{owl:complementOf} can be used as a workaround for expressing optional patterns (\texttt{OPT}) and for enabling soft negation to represent \texttt{MINUS} operations. Although OWL does not natively support the \texttt{FILTER} operator, constraints on property values can still be expressed using OWL restrictions and property assertions. However, OWL does not have a direct equivalent of \texttt{BIND}. Instead, the start time and deadline of an obligation can be directly assigned as fixed values in the policy using data property assertions or class-level restrictions.
In our GitHub repository\footnote{\url{https://github.com/Ines-Akaichi/Temporal-GUCON/tree/main/instantiation}}, we provide  instantiations of Scenario \textbf{3} using SHACL and OWL.
One avenue worth exploring is how SHACL implementations and OWL reasoners can be leveraged to carry out our defined reasoning tasks.

\emph{Usability.} This criteria concerns the aspects of the model that affect a user’s ability to define and work with it effectively. The syntax of GUCON rules is deliberately concise: both the condition and the extended action of a rule follow the SPARQL-star syntax. We have also demonstrated how this syntax can be seamlessly mapped to the Turtle language, a widely adopted and human-readable serialization. Both syntax are advantageous for Semantic Web engineers who are already familiar with existing tools. Looking ahead, we envision the development of dedicated policy editors that allow users to define policies in a more intuitive and user-friendly manner, with automatic translation into either Turtle or the GUCON syntax.

\begin{figure}[t] 
  \centering
   \includegraphics[width=0.60\linewidth]{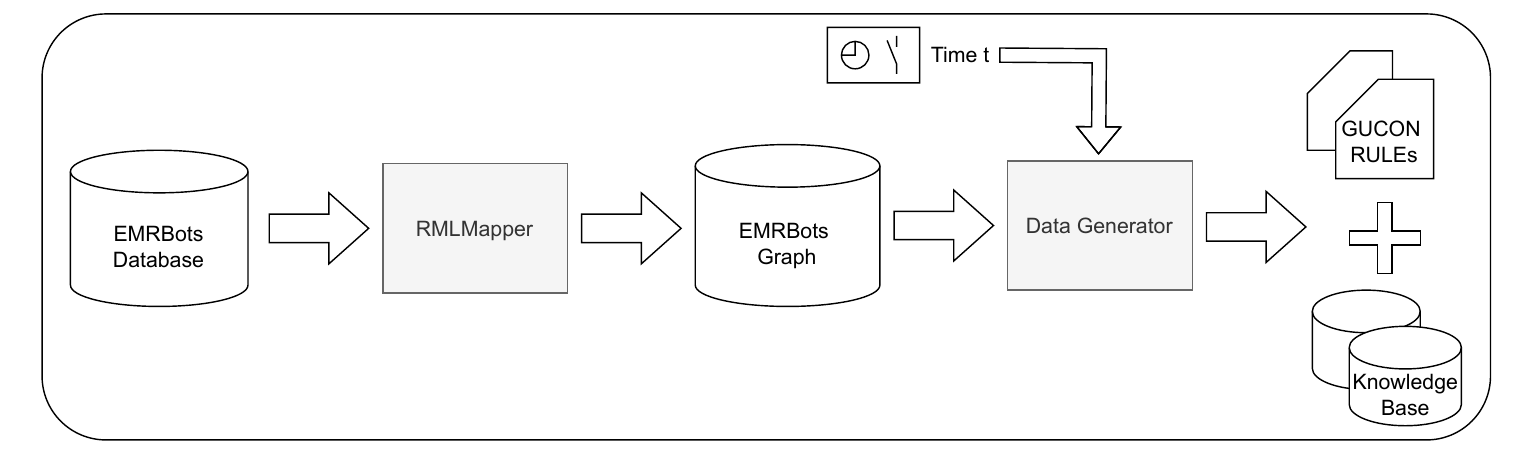}
  \captionof{figure}{Data Generation Pipeline}
  \label{fig:pipeline}
\end{figure}

\subsection{Performance \& Scalability}
To evaluate the performance and scalability of our prototype, we design our evaluation based on two key aspects: i) choosing appropriate data to feed our prototype and defining the corresponding data generation process; ii) designing benchmarking tasks based on choke point analysis. A choke-point analysis is aimed at identifying important technical challenges that should be evaluated in the context of the underlying architectural solution.
These tasks are evaluated using specific key performance indicators primarily execution time.
\subsubsection{Data Generation}
\begin{figure}[t] 
  \centering
  \includegraphics[width=0.70\linewidth]{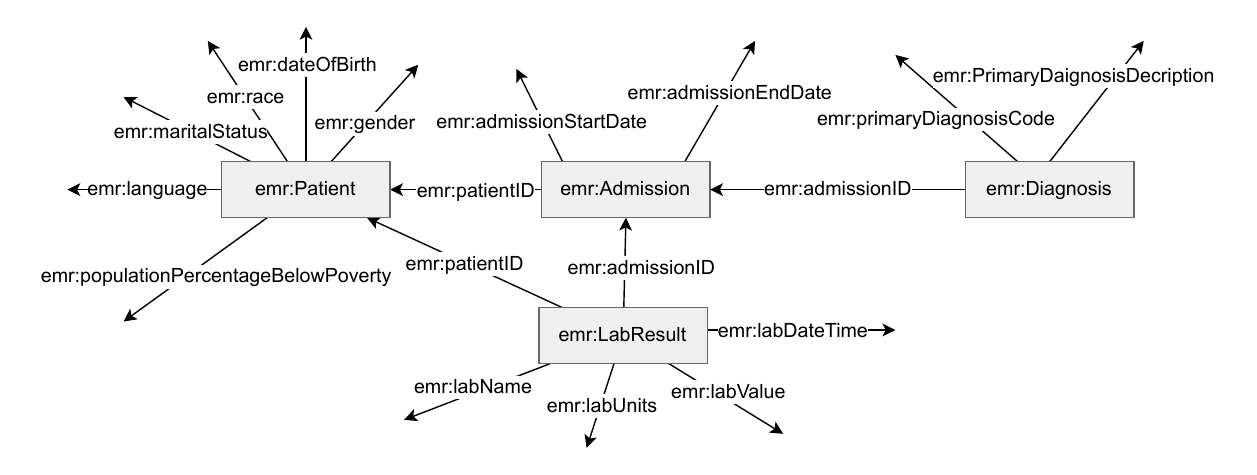}
  \captionof{figure}{The EMRBots Schema}
  \label{fig:emrbots}
\end{figure}

For our experimental evaluation, we use a dataset that closely reflects our target use case. The EMRBots dataset\footnote{EMRBots, \url{https://github.com/kartoun/emrbots}} provides synthetic electronic medical records for 100 patients. It emulates the structure and content of real-world medical databases, including patient admissions, demographics, socioeconomic information, lab results, and more. 
The EMRBots dataset encompasses records for 100 patients, including 372 admissions, 372 diagnosis reports, and a total of 110,107 lab test entries.
The RDF graph derived from the EMRBots dataset forms the basis for generating GUCON rules and constructing KBs, both of which are employed by our GUCON Obligation Manager. The data generation pipeline is illustrated in Figure~\ref{fig:pipeline}. Originally provided as a relational database in plain text files, the EMRBots dataset was transformed into an RDF graph using the RDF Mapping Language (RML)\footnote{RML, \url{https://rml.io/specs/rml/}} and the RMLMapper tool\footnote{RMLMapper, \url{https://github.com/RMLio/rmlmapper-java}}. The schema of the resulting EMRBots graph is shown in Figure~\ref{fig:emrbots}.
This RDF graph is then processed by a Data Generator, which produces GUCON rules and the associated KB. 
The rules are generated using a predefined template, as shown in Listing~\ref{lst:template}.
\begin{lstlisting}[
basicstyle=\scriptsize, %or \small or \footnotesize etc.
caption={GUCON Template Rule},captionpos=b, label={lst:template}, numbers=left,frame=single]
{?e    <random predicate>    ?r . ?e    <random predicate>    ?v .
BIND (<random startTime> as ?startTime)
BIND (<random deadline> as ?deadline)}
-> O <<?e <gucon:random action>  ?r>> gucon:startTime ?startTime  gucon:deadline ?deadline
\end{lstlisting}
The template consists of two triple patterns that incorporate two distinct predicates selected from the EMRBots graph. Up to 56 unique rules can be generated based on these predicate combinations. 
Additionally, the template uses BIND to bind the start time and the deadline of the rule to generated date-times. 
The resulting KB comprises two parts: the DKB, which consists of the EMRBots graph itself, and the AKB, which includes generated events corresponding to the action triples defined by the rules. The full KB comprises around 2400000 triples.
The Data Generator also requires an input timestamp to ensure that the temporal components of both the rules and execution traces are generated in a meaningful and coherent manner.

\subsubsection{Benchmark Tasks}
The benchmark tasks address two primary scalability choke points: increasing data volume and increasing rule count. An increase in data volume corresponds to a higher number of RDF triples in the KB, while an increase in rule count reflects a growing policy size.
Table~\ref{tab:tasks} outlines the different tasks associated with each choke point and their respective scaling strategy. Specifically, the policy size is incremented in steps of 4 rules, and the KB size is scaled in increments of 108,000 triples.

\paragraph{Selectivity.}
We first analyzed the selectivity of the generated rules against the KB and grouped them according to the number of matches they return:
\begin{enumerate}
    \item  Low selectivity: 21 rules return  300-400 matches.
    \item  Medium selectivity: 14 rules return 1116–1488 matches.
    \item  High selectivity: 21 rules return 330K–551K matches.
\end{enumerate}
From this distribution, benchmark rules were selected from one group at a time. This ensures that rules added at each scaling step exhibit comparable match counts, thereby making the benchmark more controlled and meaningful. For the reported experiments, we used rules from the high-selectivity group. Likewise, subsets of the KB were generated from the initial dataset while preserving proportional match counts for each query.

\subsection{Experimental Setup \& Result}
\begin{figure}[t] 
  \centering
  \begin{subfigure}[t]{0.48\textwidth}
    \centering
    \includegraphics[width=\linewidth]{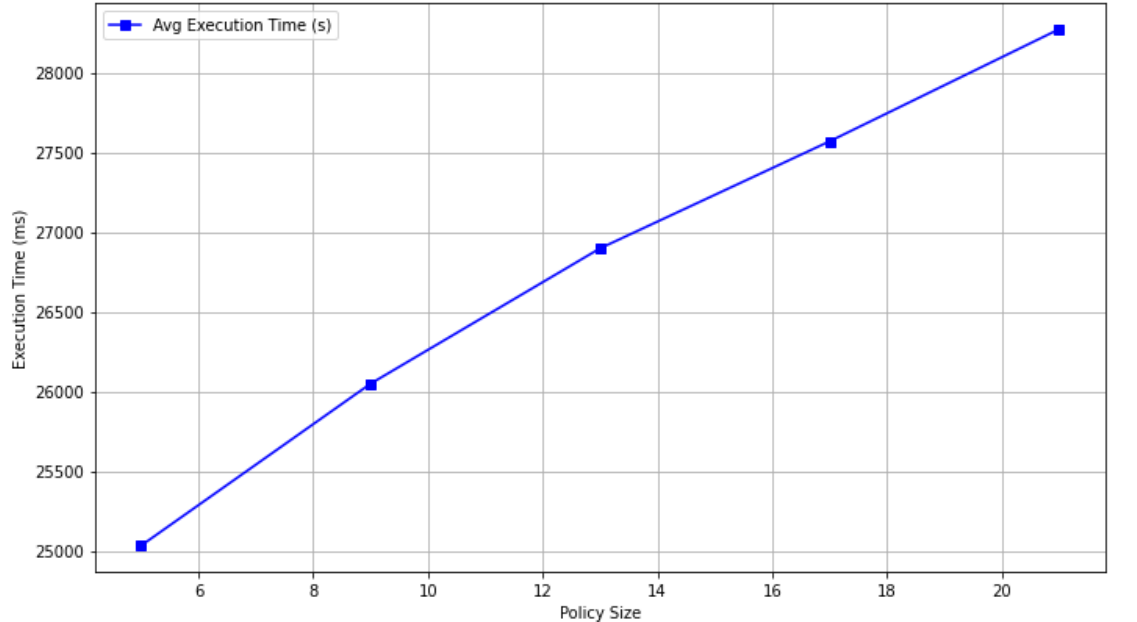}
    \caption{Task 1: Execution Time in Terms of Policy Size}
    \label{fig:task1-exec}
  \end{subfigure}
  \hfill
  \begin{subfigure}[t]{0.48\textwidth}
    \centering
     \includegraphics[width=\linewidth]{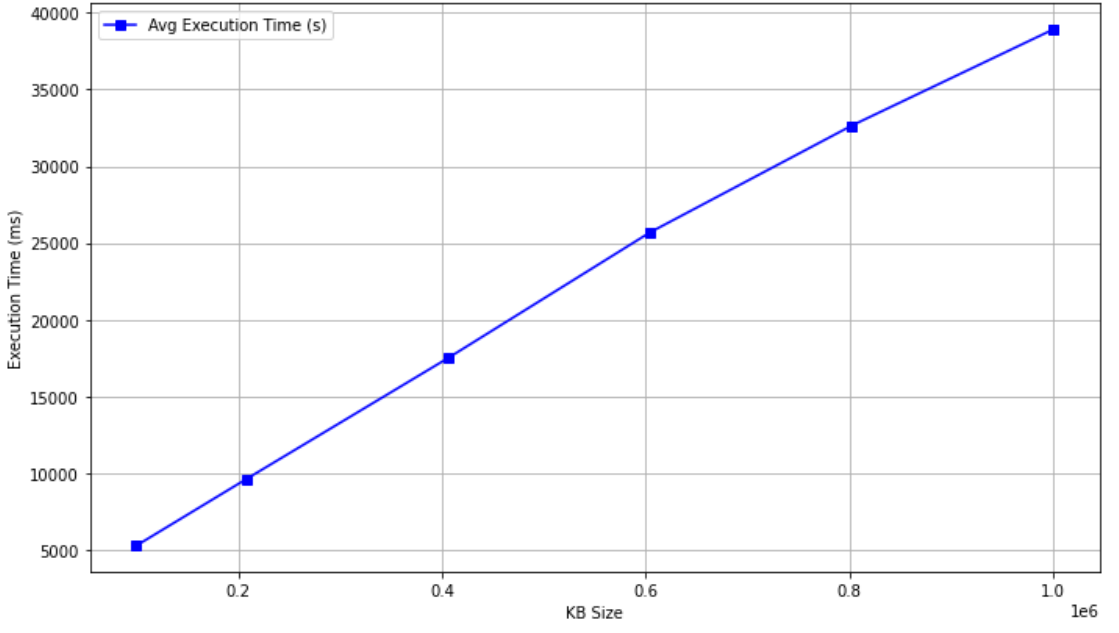}
    \caption{Task 2: Execution Time in Terms of KB Size}
    \label{fig:task2-exec}
  \end{subfigure}
  \caption{Experiments Results}
  \label{fig:results}
\end{figure}

Experiments were run on a virtual machine with 125GB of RAM, equipped with an Intel® Xeon® Silver 4114 CPU@2.20GHz, featuring 10 physical cores on Fedora Linux 42 (Adams).
The TDB2 was parameterized to use an IRI that labels the loaded KB. All the subsequent processing steps are performed in memory.
All calculations presented were based on an average of 10 response
times excluding the two slowest and fastest times  in a cold cache scenario (caches are empty at the start of each process).
Figure \ref{fig:results} presents the results of our experiments. Specifically, Figure \ref{fig:task1-exec} and  \ref{fig:task2-exec} shows a steady linear increase in execution time (in ms) as the size of policy and KB increases. This indicates that the prototype maintains predictable performance behavior with respect to larger policies and knowledge bases.


\section{Related Work}

\label{sec:relatedwork}

\begin{table}[t]
\centering
\caption{Benchmark Tasks}
\label{tab:tasks}
\begin{tabular}{llll}
\hline
Task &
  Policy &
  Knowledge Base &
  Choke Point \\
  \hline
1 &
  \begin{tabular}[c]{@{}l@{}}5\\ 9\\ 13\\ 17\\ 21\end{tabular} &
  406000 &
  Increasing number of rules \\
    \hline
2 &
  13 &
  \begin{tabular}[c]{@{}l@{}}100000\\ 208000\\ 406000\\ 604000\\ 802000\\ 1000000\end{tabular} &
  Increasing data volume
\\
\hline
\end{tabular}
\end{table}

There are several works that focus on usage control in general and obligation monitoring in particular. UCON \cite{Sandhu2004} is an abstract model that extends access control with the concepts of obligations, decision continuity and attribute mutability. 
Although several formalisms have been suggested for UCON, and attempts have been made to include it in standard representation languages \cite{LAZOUSKI201081, Maurizio2010}, there is currently no established reference or standard policy specification and implementation for UCON. As a result, UCON has not gained widespread adoption in industry.
Another language, OSL, formalized in Z \cite{Hilty2007}, is utilized to express conditional prohibitions and obligations. Although one can express temporal constraints using OSL, the obligations in OSL do not support the notion of obligation states. Additionally, it is unclear how said obligations should be enforced.

In the Semantic Web community, several general policy languages and frameworks have been proposed, including Protune \cite{protune2010}, Rei \cite{Kagal2002}, Ponder \cite{Damianou2001},  KAoS \cite{Uszok2003}, and the DSA policy framework \cite{10.1007/978-3-030-62466-8_30}. Protune focuses on access control and trust management, but lacks support for expressing obligations. While, Rei, Ponder, KAoS, and the DSA framework can express obligations, their design lacks the states of obligations, which limits their ability to monitor obligation lifecycles. Furthermore, it remains unclear how obligations expressed in these languages are enforced in practice. Additionally, Ponder lacks formal semantics, which limits its applicability in policy reasoning. 

In terms of policy standards, ODRL\footnote{ODRL, \url{https://www.w3.org/TR/odrl-model/}} is a W3C recommendation that provides a model and vocabulary for describing policies, including obligations. Although ODRL does not have yet an official formal semantics, several efforts have attempted to formalize it, either by using web ontologies \cite{GarcaFormalisingOS} or by defining operational semantics through rule-based reasoning \cite{Fornara2018, Steyskal2015TowardsFS}.
Notably, \citet{Fornara2019} focused on enriching ODRL by extending the model with the notions of permission and obligation states. The operational semantics of this extended model are implemented using a production rule system. 
Our extended GUCON model builds upon the obligation states formally defined by \citet{Fornara2019}. However, we simplify the GUCON model by including only temporal metaproperties and define the formal semantics of obligation states separately.
Currently, a W3C group\footnote{ODRL Formal Semantics, \url{https://w3c.github.io/odrl/formal-semantics/}} is working on defining the formal semantics of ODRL; which is currently only described informally in plain English, without any formal specification.
\citet{Cimmino2025} highlight several limitations of ODRL. Although ODRL supports expressing various types of constraints; such as temporal and spatial constraints; they lack formal semantics. In particular, it remains unclear how to reason effectively about obligation start times and deadlines. Furthermore, ODRL does not incorporate the notion of the state of affairs, which complicates the enforcement and/ or policy re-evaluation whenever the policy or context changes.
In contrast, the state of the affairs, represented as a KB, is a fundamental component of our GUCON framework. GUCON also leverages graph patterns, enabling straightforward and dynamic policy re-evaluation against the current state of the world.
SHACL is another standard that has been explored for expressing policies. \citet{robaldo2023compliance} propose using SHACL to represent norms and leveraging SHACL implementations for compliance checking. However, it remains unclear how these implementations can be used to reason over the states of obligations. 
Another widely adopted standard is the XACML policy language and framework \cite{XACML2013}, originally developed for managing access control policies. Several studies have extended XACML with UCON capabilities \cite{10.1007/978-1-4419-6794-7_11,kateb2014}. Another related standard is the Abbreviated Language for Authorization (ALFA)\footnote{ALFA, \url{https://alfa.guide/}}, a domain-specific language for specifying authorization policies. However, both XACML (including its extensions) and ALFA lack formal foundations.
For a more in-depth analysis of the state of the art in usage control solutions, the survey in \cite{AKAICHI2025100698} offers a comprehensive overview of the existing gaps in this field.

\section{Conclusion}
\label{sec:conclusion}


In this paper, we presented an extension of the GUCON model that integrates temporal properties into obligations, enabling their continuous monitoring and reasoning over time. By incorporating start times and deadlines, our approach supports the formal assessment of obligation states and allows the evaluation of compliance with respect to a temporal knowledge graph (state of affairs). We further demonstrated how the extended model can be instantiated using RDF-star and SPARQL-star and introduced an Obligation State Manager that tracks obligation states and checks compliance against the state of affairs. 

To assess our approach, we evaluated the extended model using Thörn’s criteria for model quality. The evaluation highlighted several strengths: (i) adaptability, as the model can be easily extended with new constraints; (ii) reusability, since the model’s generality allows it to be applied across multiple domains; (iii) formalness, due to its grounding in formal semantics; (iv) correctness, as demonstrated through the use case and corresponding test scenarios;  (v) mobility, by showing different ways of instantiating the model using Semantic Web languages; and  (vi) usability, as GUCON deliberately employs concise syntax to reduce verbosity.
We also evaluated the performance and scalability of the Obligation State Manager prototype, with results showing that the system scales well with increasing numbers of obligation policies and larger knowledge graphs.

Future work will focus on extending the model with additional constraints, such as spatial properties, and investigating the interplay between temporal and spatial aspects for capturing realistic usage control scenarios. To strengthen interoperability, we aim to develop systematic mappings between different instantiations of GUCON, such as OWL and SHACL. Finally, we plan to address usability by designing intuitive interfaces that enable users to craft and manage GUCON policies more effectively.

\begin{acks}
This work is funded by the European Union Horizon 2020 research and innovation programme under the Marie Skłodowska-Curie grant agreement No 860801 and the FWF Austrian Science Fund 
[10.55776/COE12]. Sabrina Kirrane is also funded by the FWF Austrian Science Fund and the Internet Foundation Austria under the FWF Elise Richter and netidee SCIENCE programmes as project number V 759-N.
\end{acks}





\nocite{*}
\bibliographystyle{plainnat} 
\bibliography{bibliography}        

\begin{thebibliography}{50}
\providecommand{\natexlab}[1]{#1}
\providecommand{\url}[1]{\texttt{#1}}
\expandafter\ifx\csname urlstyle\endcsname\relax
  \providecommand{\doi}[1]{doi: #1}\else
  \providecommand{\doi}{doi: \begingroup \urlstyle{rm}\Url}\fi

\bibitem[Akaichi and Kirrane(2025)]{AKAICHI2025100698}
Ines Akaichi and Sabrina Kirrane.
\newblock A comprehensive review of usage control frameworks.
\newblock \emph{Computer Science Review}, 56, 2025.

\bibitem[Akaichi et~al.(2023)Akaichi, Flouris, Fundulaki, and Kirrane]{akaichi2023}
Ines Akaichi, Giorgos Flouris, Irini Fundulaki, and Sabrina Kirrane.
\newblock Gucon: A generic graph pattern based policy framework for usage control enforcement.
\newblock In \emph{Rules and Reasoning: 7th International Joint Conference, RuleML+RR 2023, Oslo, Norway, September 18–20, 2023, Proceedings}, 2023.

\bibitem[Allen(1983)]{Allen1983}
James~F. Allen.
\newblock Maintaining knowledge about temporal intervals.
\newblock \emph{Communications of the ACM}, 26, 1983.

\bibitem[Asma et~al.(2024)Asma, Hernández, Galárraga, Flouris, Fundulaki, and Hose]{Asma2024}
Zubaria Asma, Daniel Hernández, Luis Galárraga, Giorgos Flouris, Irini Fundulaki, and Katja Hose.
\newblock Npcs: Native provenance computation for sparql.
\newblock \emph{WWW 2024 - Proceedings of the ACM Web Conference}, 2024.

\bibitem[Bonatti et~al.(2010)Bonatti, De~Coi, Olmedilla, and Sauro]{protune2010}
Piero Bonatti, J.L. De~Coi, Daniel Olmedilla, and Luigi Sauro.
\newblock A rule-based trust negotiation system.
\newblock \emph{IEEE Transactions on Knowledge and Data Engineering}, 22\penalty0 (11), 2010.

\bibitem[Booshehri et~al.(2021)Booshehri, Emele, Flügel, Förster, Frey, Frey, Glauer, Hastings, Hofmann, Hoyer-Klick, Hülk, Kleinau, Knosala, Kotzur, Kuckertz, Mossakowski, Muschner, Neuhaus, Pehl, Robinius, Sehn, and Stappel]{BOOSHEHRI2021100074}
Meisam Booshehri, Lukas Emele, Simon Flügel, Hannah Förster, Johannes Frey, Ulrich Frey, Martin Glauer, Janna Hastings, Christian Hofmann, Carsten Hoyer-Klick, Ludwig Hülk, Anna Kleinau, Kevin Knosala, Leander Kotzur, Patrick Kuckertz, Till Mossakowski, Christoph Muschner, Fabian Neuhaus, Michaja Pehl, Martin Robinius, Vera Sehn, and Mirjam Stappel.
\newblock Introducing the open energy ontology: Enhancing data interpretation and interfacing in energy systems analysis.
\newblock \emph{Energy and AI}, 5, 2021.

\bibitem[Bradshaw et~al.(1997)Bradshaw, Dutfield, Benoit, and Woolley]{Bradshaw1997}
Jm~Jeffrey~M Bradshaw, Stewart Dutfield, Pete Benoit, and John D~Jd Woolley.
\newblock Kaos: toward an industrial-strength open agent architecture.
\newblock \emph{Software Agents}, 1997.

\bibitem[Cimmino and Fornara(2025)]{Cimmino2025}
Andrea Cimmino and Nicoletta Fornara.
\newblock Improving odrl 2.2: current limitations and theoretical solutions.
\newblock In \emph{Joint Proceedings of the ESWC 2025 Workshops and Tutorials co-located with 22nd Extended Semantic Web Conference (ESWC 2025)}, 2025.

\bibitem[Cimmino et~al.(2025)Cimmino, Cano-Benito, and Garc\'{\i}a-Castro]{Cimmino2024}
Andrea Cimmino, Juan Cano-Benito, and Ra\'{u}l Garc\'{\i}a-Castro.
\newblock Open digital rights enforcement framework (odre): From descriptive to enforceable policies.
\newblock \emph{Comput. Secur.}, 150, 2025.

\bibitem[Colombo et~al.(2010{\natexlab{a}})Colombo, Lazouski, Martinelli, and Mori]{10.1007/978-1-4419-6794-7_11}
Maurizio Colombo, Aliaksandr Lazouski, Fabio Martinelli, and Paolo Mori.
\newblock A proposal on enhancing xacml with continuous usage control features.
\newblock In \emph{Grids, P2P and Services Computing}, 2010{\natexlab{a}}.

\bibitem[Colombo et~al.(2010{\natexlab{b}})Colombo, Lazouski, Martinelli, and Mori]{Maurizio2010}
Maurizio Colombo, Aliaksandr Lazouski, Fabio Martinelli, and Paolo Mori.
\newblock A proposal on enhancing xacml with continuous usage control features.
\newblock In \emph{Grids, P2P and Services Computing}, 2010{\natexlab{b}}.

\bibitem[Damianou et~al.(2001)Damianou, Dulay, Lupu, and Sloman]{Damianou2001}
Nicodemos Damianou, Naranker Dulay, Emil Lupu, and Morris Sloman.
\newblock The ponder policy specification language.
\newblock \emph{Lecture Notes in Computer Science (including subseries Lecture Notes in Artificial Intelligence and Lecture Notes in Bioinformatics)}, 1995, 2001.

\bibitem[De~Vos et~al.(2019)De~Vos, Kirrane, Padget, and Satoh]{DeVos2019}
Marina De~Vos, Sabrina Kirrane, Julian Padget, and Ken Satoh.
\newblock Odrl policy modelling and compliance checking.
\newblock In \emph{Rules and Reasoning: Third International Joint Conference, RuleML+RR 2019, Bolzano, Italy, September 16–19, 2019, Proceedings}, 2019.

\bibitem[Dimishkovska(2017)]{Dimishkovska2017}
Ana Dimishkovska.
\newblock Deontic logic and legal rules.
\newblock \emph{Encyclopedia of the Philosophy of Law and Social Philosophy}, 2017.

\bibitem[{Erik Rissanen}(2013)]{XACML2013}
{Erik Rissanen}.
\newblock extensible access control markup language (xacml) version 3.0., 2013.
\newblock URL \url{http://docs.oasis-open.org/ xacml/3.0/xacml-3.0-core-spec-os-en.pdf}.
\newblock Online; accessed 14 February 2023.

\bibitem[Fernández et~al.(2019)Fernández, Bonatti, Milosevic, and Langens]{Javier2019}
Javier~D. Fernández, P.A. Bonatti, U.~Milosevic, and Jonathan Langens.
\newblock {Scalability and Robustness testing report V2}.
\newblock Technical report, H2020 Project, 2019.
\newblock URL \url{https://specialprivacy.ercim.eu/images/documents/SPECIAL_D35_M27_V10.pdf}.

\bibitem[Fornara and Colombetti(2018)]{Fornara2018}
Nicoletta Fornara and Marco Colombetti.
\newblock Operational semantics of an extension of odrl able to express obligations.
\newblock \emph{Lecture Notes in Computer Science (including subseries Lecture Notes in Artificial Intelligence and Lecture Notes in Bioinformatics)}, 10767 LNAI, 2018.

\bibitem[Fornara et~al.(2019)Fornara, Chiappa, and Colombetti]{Fornara2019}
Nicoletta Fornara, Alessia Chiappa, and Marco Colombetti.
\newblock Using semantic web technologies and production rules for reasoning on obligations and permissions.
\newblock In Marin Lujak, editor, \emph{Agreement Technologies}, 2019.

\bibitem[Galton(1991)]{Galton1991}
Antony Galton.
\newblock Reified temporal theories and how to unreify them.
\newblock \emph{12th International Joint Conference on Artificial Intelligence}, 88, 1991.

\bibitem[Garc{\'\i}a et~al.(2005)Garc{\'\i}a, Gil, Gallego, and Delgado]{GarcaFormalisingOS}
Roberto Garc{\'\i}a, Rosa Gil, Isabel Gallego, and Jaime Delgado.
\newblock Formalising odrl semantics using web ontologies.
\newblock In \emph{Proc. 2nd Intl. ODRL Workshop}, 2005.

\bibitem[Gutierrez et~al.(2005)Gutierrez, Hurtado, and Vaisman]{10.1007/11431053_7}
Claudio Gutierrez, Carlos Hurtado, and Alejandro Vaisman.
\newblock Temporal rdf.
\newblock In Asunci{\'o}n G{\'o}mez-P{\'e}rez and J{\'e}r{\^o}me Euzenat, editors, \emph{The Semantic Web: Research and Applications}, 2005.

\bibitem[Gutierrez et~al.(2007)Gutierrez, Hurtado, and Vaisman]{4039284}
Claudio Gutierrez, Carlos~A. Hurtado, and Alejandro Vaisman.
\newblock Introducing time into rdf.
\newblock \emph{IEEE Transactions on Knowledge and Data Engineering}, 19\penalty0 (2), 2007.

\bibitem[Hilty et~al.(2007)Hilty, Pretschner, Basin, Schaefer, and Walter]{Hilty2007}
M.~Hilty, A.~Pretschner, D.~Basin, C.~Schaefer, and T.~Walter.
\newblock A policy language for distributed usage control.
\newblock In Joachim Biskup and Javier L{\'o}pez, editors, \emph{Computer Security -- ESORICS 2007}, 2007.

\bibitem[Hobbs and Pan(2004)]{Hobbs2004}
Jerry~R. Hobbs and Feng Pan.
\newblock An ontology of time for the semantic web.
\newblock \emph{ACM Transactions on Asian Language Information Processing}, 3, 2004.

\bibitem[Hoffart et~al.(2013)Hoffart, Suchanek, Berberich, and Weikum]{Hoffart2013}
Johannes Hoffart, Fabian~M. Suchanek, Klaus Berberich, and Gerhard Weikum.
\newblock Yago2: A spatially and temporally enhanced knowledge base from wikipedia.
\newblock \emph{Artificial Intelligence}, 194, 2013.

\bibitem[Hu et~al.(2011)Hu, Wang, She, and Wang]{10.1007/978-3-642-18333-1_18}
Siquan Hu, Haiou Wang, Chundong She, and Junfeng Wang.
\newblock Agont: Ontology for agriculture internet of things.
\newblock In Daoliang Li, Yande Liu, and Yingyi Chen, editors, \emph{Computer and Computing Technologies in Agriculture IV}, 2011.

\bibitem[{Iannella, Renato and Villata, Serena}(2018)]{ODRL2018}
{Iannella, Renato and Villata, Serena}.
\newblock Odrl information model 2.2, 2018.
\newblock URL \url{https://www.w3.org/TR/odrl-model/}.
\newblock Online; accessed 28 April 2025.

\bibitem[Kagal and Berners-Lee(2005)]{Kagal2005}
L.~Kagal and T.~Berners-Lee.
\newblock Trein: Where policies meet rules in the semantic web.
\newblock Technical report, CSAIL, Massachusetts Institute of Technology, 2005.
\newblock URL \url{http://groups.csail.mit.edu/dig/2005/05/rein/rein-paper.pdf.}

\bibitem[Kagal(2002)]{Kagal2002}
Lalana Kagal.
\newblock Rei 1 : A policy language for the me-centric project rei £ : A policy language for the me-centric project.
\newblock Technical report, HP Laboratories, 2002.
\newblock URL \url{https://ebiquity.umbc.edu/_file_directory_/papers/57.pdf}.

\bibitem[Kateb et~al.(2014)Kateb, Elrakaiby, Mouelhi, Rubab, and Le~Traon]{kateb2014}
Donia Kateb, Yehia Elrakaiby, Tejeddine Mouelhi, Iram Rubab, and Yves Le~Traon.
\newblock Towards a full support of obligations in xacml.
\newblock In \emph{Risks and Security of Internet and Systems}, 08 2014.

\bibitem[Katsumi and Fox(2018)]{KATSUMI201853}
Megan Katsumi and Mark Fox.
\newblock Ontologies for transportation research: A survey.
\newblock \emph{Transportation Research Part C: Emerging Technologies}, 89, 2018.

\bibitem[Knez and Žitnik(2023)]{Knez2023}
Timotej Knez and Slavko Žitnik.
\newblock Event-centric temporal knowledge graph construction: A survey.
\newblock \emph{Mathematics}, 11, 12 2023.

\bibitem[Lazouski et~al.(2010)Lazouski, Martinelli, and Mori]{LAZOUSKI201081}
Aliaksandr Lazouski, Fabio Martinelli, and Paolo Mori.
\newblock Usage control in computer security: A survey.
\newblock \emph{Computer Science Review}, 4\penalty0 (2), 2010.

\bibitem[Park and Sandhu(2004)]{Sandhu2004}
Jaehong Park and Ravi Sandhu.
\newblock The uconabc usage control model.
\newblock \emph{ACM Transactions on Information and System Security}, 7, 2004.

\bibitem[P{\'e}rez et~al.(2006)P{\'e}rez, Arenas, and Gutierrez]{Perez2006}
Jorge P{\'e}rez, Marcelo Arenas, and Claudio Gutierrez.
\newblock Semantics and complexity of sparql.
\newblock In \emph{The Semantic Web - ISWC 2006}, 2006.

\bibitem[Piryani et~al.(2023)Piryani, Aussenac-Gilles, and Hernandez]{Piryani2023}
Rajesh Piryani, Nathalie Aussenac-Gilles, and Nathalie Hernandez.
\newblock Comprehensive survey on ontologies about event.
\newblock \emph{CEUR Workshop Proceedings}, 3443, 2023.

\bibitem[Pretschner et~al.(2008)Pretschner, Hilty, Sch{\"{u}}tz, Schaefer, and Walter]{DBLP:journals/ieeesp/PretschnerHSSW08}
Alexander Pretschner, Manuel Hilty, Florian Sch{\"{u}}tz, Christian Schaefer, and Thomas Walter.
\newblock Usage control enforcement: Present and future.
\newblock \emph{{IEEE} Security \& Privacy}, 6\penalty0 (4), 2008.

\bibitem[Prud'hommeaux and Seaborne(2008)]{Prudhommeaux2006}
Eric Prud'hommeaux and Andy Seaborne.
\newblock {SPARQL Query Language for RDF}.
\newblock \url{https://www.w3.org/TR/rdf-sparql-query/}, 2008.
\newblock W3C Recommendation 15 January 2008.

\bibitem[Robaldo et~al.(2023)Robaldo, Batsakis, Calegari, and et~al.]{robaldo2023compliance}
L.~Robaldo, S.~Batsakis, R.~Calegari, and et~al.
\newblock Compliance checking on first-order knowledge with conflicting and compensatory norms: a comparison among currently available technologies.
\newblock \emph{Artificial Intelligence and Law}, 2023.

\bibitem[Robaldo(2021{\natexlab{a}})]{Robaldo2021}
Livio Robaldo.
\newblock Towards compliance checking in reified i/o logic via shacl.
\newblock In \emph{Proceedings of the Eighteenth International Conference on Artificial Intelligence and Law}, 2021{\natexlab{a}}.

\bibitem[Robaldo(2021{\natexlab{b}})]{Robaldo2021a}
Livio Robaldo.
\newblock Towards compliance checking in reified i/o logic via shacl.
\newblock \emph{Proceedings of the 18th International Conference on Artificial Intelligence and Law, ICAIL 2021}, 2021{\natexlab{b}}.

\bibitem[Robaldo and Sun(2017)]{Robaldo2017}
Livio Robaldo and Xin Sun.
\newblock Reified input/output logic: Combining input/output logic and reification to represent norms coming from existing legislation.
\newblock \emph{Journal of Logic and Computation}, 27, 2017.

\bibitem[Röder et~al.(2020)Röder, Kuchelev, and Ngomo]{doi:10.3233/DS-190021}
Michael Röder, Denis Kuchelev, and Axel-Cyrille~Ngonga Ngomo.
\newblock Hobbit: A platform for benchmarking big linked data.
\newblock \emph{Data Science}, 3\penalty0 (1), 2020.

\bibitem[Santos et~al.(2020)Santos, Mulvehill, Erickson, McCusker, Gordon, Xie, Stouffer, Capraro, Pidwerbetsky, Burgess, Berlinsky, Turck, Ashdown, and McGuinness]{10.1007/978-3-030-62466-8_30}
Henrique Santos, Alice Mulvehill, John~S. Erickson, Jamie~P. McCusker, Minor Gordon, Owen Xie, Samuel Stouffer, Gerard Capraro, Alex Pidwerbetsky, John Burgess, Allan Berlinsky, Kurt Turck, Jonathan Ashdown, and Deborah~L. McGuinness.
\newblock A semantic framework for enabling radio spectrum policy management and evaluation.
\newblock In \emph{The Semantic Web – ISWC 2020: 19th International Semantic Web Conference, Athens, Greece, November 2–6, 2020, Proceedings, Part II}, 2020.

\bibitem[Santos et~al.(2024)Santos, Mccusker, Erickson, Mulvehill, Seneviratne, and Mcguinness]{Santos2024}
Henrique Santos, Jamie~P Mccusker, John~S Erickson, Alice~M Mulvehill, Oshani Seneviratne, and Deborah~L Mcguinness.
\newblock Towards computable and explainable policies using semantic web standards.
\newblock In \emph{15th Workshop on Ontology Design and Patterns co-located with ISWC 2024}, 2024.

\bibitem[Steyskal and Polleres(2015)]{Steyskal2015TowardsFS}
Simon Steyskal and Axel Polleres.
\newblock Towards formal semantics for odrl policies.
\newblock In \emph{International Web Rule Symposium}, 2015.

\bibitem[Thörn(2010)]{thorn2010quality}
Christer Thörn.
\newblock \emph{On the Quality of Feature Models}.
\newblock Ph.d. thesis, Linköping University, 2010.
\newblock URL \url{https://liu.diva-portal.org/smash/get/diva2:313940/FULLTEXT01.pdf}.

\bibitem[Toninelli et~al.(2005)Toninelli, Bradshaw, Kagal, Montanari, et~al.]{Toninelli2005}
Alessandra Toninelli, Jeffrey Bradshaw, Lalana Kagal, Rebecca Montanari, et~al.
\newblock Rule-based and ontology-based policies : Toward a hybrid approach to control agents in pervasive environments.
\newblock \emph{Informatica}, 2005.

\bibitem[Uszok et~al.(2003)Uszok, Bradshaw, Jeffers, Suri, Hayes, Breedy, Bunch, Johnson, Kulkarni, and Lott]{Uszok2003}
A.~Uszok, J.~Bradshaw, R.~Jeffers, N.~Suri, P.~Hayes, M.~Breedy, L.~Bunch, M.~Johnson, S.~Kulkarni, and J.~Lott.
\newblock Kaos policy and domain services: Toward a description-logic approach to policy representation, deconfliction, and enforcement.
\newblock \emph{Proceedings - POLICY 2003: IEEE 4th International Workshop on Policies for Distributed Systems and Networks}, 2003.

\bibitem[Vila and Reichgelt(1996)]{Vila1996}
Lluís Vila and Han Reichgelt.
\newblock The token reification approach to temporal reasoning.
\newblock \emph{Artificial Intelligence}, 83, 1996.

\end{thebibliography}

%


\end{document}